\documentclass[lettersize,journal]{IEEEtran}
\usepackage{amsmath,amsfonts}
\usepackage{algorithmic}
\usepackage{algorithm}
\usepackage{array}
\usepackage[caption=false,font=normalsize,labelfont=sf,textfont=sf]{subfig}
\usepackage{textcomp}
\usepackage{stfloats}
\usepackage{url}
\usepackage{verbatim}
\usepackage{graphicx}
\usepackage{cite}
\usepackage{color, xcolor}
\usepackage{placeins}
\usepackage{siunitx}
\usepackage{enumitem}
\usepackage{booktabs}
\usepackage{siunitx}
\begin{document}

\title{Robust Detection of Underwater Target Against  \\ Non-Uniform Noise With Optical Fiber DAS Array}

\author{Siyuan Cang\textsuperscript{\textdagger}, \IEEEmembership{Member, IEEE}, Cong Liu\textsuperscript{\textdagger}, Xueli Sheng, Xiaoming Cui, Chao Li, Changxin Fa, Jiantong Chen, Chaoran Yang, \IEEEmembership{Member, IEEE}, and Huayong Yang
	\thanks{Siyuan Cang, Xiaoming Cui, Changxin Fa, Jiantong Chen, Chaoran Yang and Huayong Yang are with Southern Marine Science and Engineering Guangdong Laboratory (Guangzhou), Guangzhou 511458, China (e-mail: cangsiyuan@gmlab.ac.cn, cuixm@gmlab.ac.cn, fachangxin@hrbeu.edu.cn, chen\_jiantong0927@163.com, yang\_chaoran@gmlab.ac.cn, yanghy@gmlab.ac.cn).}
	\thanks{Cong Liu and Xueli Sheng are with the National Key Laboratory of Underwater Acoustic Technology, Key Laboratory of Marine Information Acquisition and Security, College of Underwater Acoustic Engineering, Harbin Engineering University, Harbin 150001, China (e-mail: m17733015243@163.com, shengxueli@hrbeu.edu.cn).}
	\thanks{Chao Li is with the Southern Marine Science and Engineering Guangdong Laboratory (Guangzhou), Guangzhou 511458, China, and also with the Ocean College, Zhejiang University, Zhoushan 316021, China (e-mail: lichao@gmlab.ac.cn).}
	\thanks{\textsuperscript{\textdagger} Siyuan Cang and Cong Liu contributed equally to this work.}	
} 

\markboth{Journal of \LaTeX\ Class Files,~Vol.~xx, No.~xx, Nov~2025}%
{Shell \MakeLowercase{\textit{et al.}}: A Sample Article Using IEEEtran.cls for IEEE Journals}


\maketitle

\begin{abstract}
The detection of underwater targets is severely affected by the non-uniform spatial characteristics of marine environmental noise. Additionally, the presence of both natural and anthropogenic acoustic sources, including shipping traffic, marine life, and geological activity, further complicates the underwater acoustic landscape. Addressing these challenges requires advanced underwater sensors and robust signal processing techniques. In this paper, we present a novel approach that leverages an optical fiber distributed acoustic sensing (DAS) system combined with a broadband generalized sparse covariance-fitting framework for underwater target direction sensing, particularly focusing on robustness against non-uniform noise. The DAS system incorporates a newly developed spiral-sensitized optical cable, which significantly improves sensitivity compared to conventional submarine cables. This innovative design enables the system to capture acoustic signals with greater precision. Notably, the sensitivity of the spiral-wound sensitized cable is around -145.69 dB $ \emph{re:} \si{1\radian / (\micro\pascal \cdot \meter)} $, as measured inside the standing-wave tube. Employing simulations, we assess the performance of the algorithm across diverse noise levels and target configurations, consistently revealing higher accuracy and reduced background noise compared to conventional beamforming techniques and other sparse techniques. In a controlled pool experiment, the correlation coefficient between waveforms acquired by the DAS system and a standard hydrophone reached 0.973, indicating high fidelity in signal capture. Moreover, during a lake field experiment, the DAS system equipped with a sensitized optical cable was deployed on the lakebed to consistently track a non-cooperative speedboat. These results confirm the effectiveness of the proposed approach in practical scenarios involving complex, non-uniform noise environments. 
\end{abstract}


\section{Introduction}
\IEEEPARstart{U}{nderwater} remote sensing systems are of notable importance in numerous applications, such as marine resource exploration~\cite{luo2014challenges,pan2023underwater}, target detection~\cite{gerg2024deep}, localization~\cite{liu2024continuous}, navigation~\cite{mikhalevsky2020deep}, and environmental monitoring~\cite{xu2019internet}. More commonly used hydrophones include conventional PVDF- and PZT-based devices as well as fiber-optic hydrophones (FOHs), which markedly improve adaptability to complex curved surfaces and open new possibilities for weak signal detection\cite{guang2024sensitivity, wang2021large}. Regrettably, the inherent complexity of the underwater environments poses notable challenges to the effectiveness of such systems~\cite{heshmati2020robust}, especially as the propagation speed of the used acoustic wavefront is affected by time-varying and often unknown factors including currents, depths, temperature, and salinity~\cite{colosi2020observations}.
\begin{figure}[!t]
	\centering
	\includegraphics[width=\columnwidth]{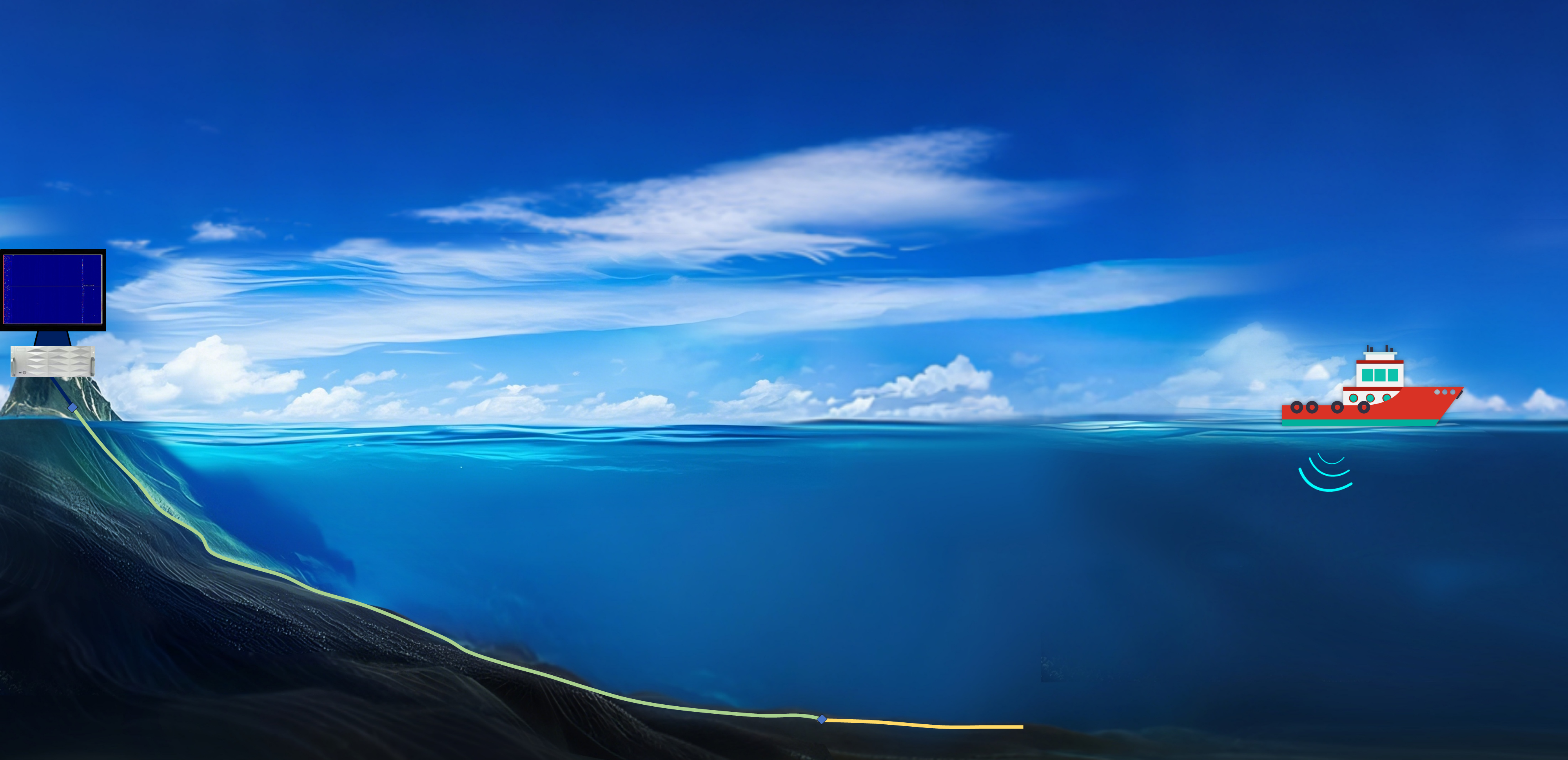}
	\caption{Schematic illustration of DAS-based underwater acoustic sensing.}
	\label{Fig:cros_Fig}
	\vspace{-3mm}
\end{figure}

To enable reliable localization of underwater targets or sound sources, one typically employs multiple sensors to determine the direction of arrival (DOA) of the acoustic wavefronts~\cite{liu2023robust,chen2024method}. Previous studies primarily addressed performance in ideal environments (e.g., signal sources are assumed to be far-field and narrowband, allowing wavefronts to be approximated as plane waves. The ambient noise field is modeled as isotropic, implying uniform noise power from all directions.) and ideal hardware configurations (e.g., high-sensitivity sensors, precisely calibrated circuits, and standard array geometries). Certain methods are employed to enhance the accuracy of estimated parameters and closely approach the Cramér-Rao lower bound (a fundamental performance metric in estimation theory, a benchmark for assessing the theoretical accuracy of signal processing) \cite{fu2019robust}. Such approaches demonstrate a strong capability in radar detection area\cite{tuo2023radar,yang2024sparse} due to the relatively stable nature of the air environment compared to underwater conditions. 

Many of these ideas and methods may also be applied for sonar signals, but the dynamic nature of the underwater acoustic channel, the anisotropic noise in the ocean environment typically reduce their effectiveness~\cite{qiu2021directional}. In particular, the presence spatially non-uniform ambient noise from both human-made and natural sources can substantially reduce the performance of conventional localization techniques
~\cite{Guo2023Variational}.

Even so, many classical methods, such as delay-and-sum beamforming, as well as subspace-based techniques like MUltiple SIgnal Classification (MUSIC) and root-MUSIC have frequently been utilized for underwater target detection~\cite{schmidt1986multiple,Lim2016GMUSIC,vallet2015performance,ahmed2021performance,Krim1996two,pesavento2023three,ghafoor2019overview,Men2024Joint,lu2024underwater}. Notably, the asymptotic ambient noise elimination based MUSIC approach was introduced to allow for the non-white ambient noise~\cite{liu2023robust}. Similarly, a series of generalized MUSIC-type methods have been introduced to allow for the typically correlated noise in underwater environments~\cite{Lim2016GMUSIC}. Although, as with many other efforts, these works focus on uniform spatial noise, and typically assumes the use of narrowband signals.

It is worth stressing that, in cases with non-uniform spatial noise, the noise covariance will generally have unequal diagonal elements and, potentially, even non-zero off-diagonal elements, preventing the use of traditional subspace approaches. As an alternative, there has been notable recent interest in formulating sparse reconstruction techniques that allow for robustness against non-uniform noise. For example, Ref. \cite{liu2022robustVBI} investigates DOA estimation with co-prime sparse array structures, where sensors are placed in a non-uniform configuration to form a larger virtual aperture and provide more degrees of freedom with fewer physical elements. In that work, a variational Bayesian inference framework was employed to achieve outlier-resistant DOA estimation, thereby reducing hardware complexity while maintaining competitive estimation accuracy. These approaches leverage the inherent sparsity of target signals in certain domains to enhance detection performance, even with fewer sensors.

To summarize, the studies mentioned concentrate on localizing, positioning, and detecting underwater targets by employing conventional hydrophone array systems that commonly incorporate piezoelectric transducer elements. However, these systems exhibit several inherent limitations, including high manufacturing and deployment costs, intensive maintenance demands, and significant difficulties in large-scale implementation\cite{lan2020acoustical, baron2020hydrophone}. These technical constraints highlight the need for more robust and cost-effective alternatives in underwater acoustic detection systems\cite{hu2020decentralized}.

In recent years, there has been a notable shift in the advancement and utilization of optical fiber Distributed Acoustic Sensing (DAS) technology\cite{he2021optical}. In contrast, DAS enables continuous monitoring over extensive oceanic regions, offering several advantages such as enhanced spatial resolution, lower operational costs, and adaptability to diverse environmental conditions\cite{landro2022sensing}. Owing to the inherent flexibility of fiber optics, DAS systems can be deployed in various configurations to support diverse ocean observation missions.

By utilizing phase-sensitive optical time-domain reflectometry ($\varphi$-OTDR), DAS system has shown considerable promise in detecting acoustic signals in shallow water environments \cite{lu2021distributed, bouffaut2022eavesdropping, landro2022sensing, douglass2023distributed, wilcock2023distributed, cang2025deploying,liu2024voiceprint, shao2025tracking, chen2025near, matsumoto2021detection}. Lu et al. pioneered a distributed optical fiber hydrophone that achieved high sensitivity and offered flexible array configurations for marine acoustic detection; however, challenges such as noise and bandwidth limitations remain \cite{lu2021distributed}. Recent studies have highlighted the potential of DAS for monitoring marine life; for instance, Arctic submarine cables were repurposed to track baleen whale vocalizations over a 120 km range, enabling three-dimensional localization and passive seismic imaging\cite{bouffaut2022eavesdropping}. Our team conducted a two-year continuous field trial utilizing the DAS system at the Xinfengjiang Reservoir in China, as illustrated in Fig.\ref{Fig:cros_Fig}, demonstrating the system's potential for integrated seismic and underwater acoustic monitoring \cite{cang2025deploying}. For ship detection, Liu et al. incorporated adaptive phase correction to achieve high-fidelity voiceprint analysis \cite{liu2024voiceprint}, while Shao et al. applied Doppler shift analysis of DAS data to track ship trajectories \cite{shao2025tracking}. Recent advances in array signal processing have significantly enhanced localization accuracy and noise suppression. Chen et al. achieved near-field multisource localization by employing near field-MUSIC beamforming techniques \cite{chen2025near}. The spiral sensitized cable is another key component for enhancing acoustic sensitivity in distributed optical fiber hydrophones (DOFH). It features a sound-sensitive mandrel wrapped with tightly wound optical fiber, protected by a permeable sheath. This design significantly improves acoustic pressure sensitivity to approximately -146 dB $ \emph{re:} \si{1\radian / (\micro\pascal \cdot \meter)} $\cite{jiajing2019distributed}, enabling high-accuracy underwater sound detection, azimuth estimation, and target tracking in $\varphi$-OTDR based DAS systems \cite{li2014fiber, zhou2015distributed, fouda2025phase}.

Despite progress, several limitations persist. Sensitivity to environmental noise and cable coupling variability continues to affect system reliability \cite{lu2021distributed}. The substantial data volume (up to 7 TB/day) demands advanced processing frameworks \cite{landro2022sensing}. Apparently, the study mentioned above does not address the heteroscedastic noise characteristics of real underwater environments or validate signal acquisition through controlled tank experiments for this new sensor.

Meanwhile, several sparse covariance-fitting methods have been introduced, building upon the seminal SParse Iterative Covariance-based Estimation (SPICE) algorithm \cite{stoica2010spice}. The SPICE was later generalized into $q$-SPICE to enhance its sparsity, robustness, and the accuracy of noise level estimates\cite{sward2018generalized}. The SPICE framework primarily targets a range of applications, such as spectrum sensing, time delay estimation, radar, and MIMO communication ~(see, e.g., \cite{zhang2018wideband, liu2024grant, zhang2021online}). To the best of our knowledge, however, the $q$-SPICE framework has not been extended to accommodate the typically broadband signals found in sonar applications. Furthermore, there have been no reported applications of this framework to underwater target detection using DAS system.

To allow the use of the $q$-SPICE framework for sonar data, we present a novel system and approach that leverage an optical fiber DAS system combined with a broadband generalized sparse covariance-fitting framework for underwater target bearing sensing, particularly focusing on robustness against non-uniform noise. The major contributions of this article can be described as follows.
\begin{itemize}
\item[(i)] {Our DAS system incorporates a newly developed spiral sensitized optical cable, which significantly enhances sensitivity compared to traditional submarine cables that have already been explored for target detection in previous studies\cite{matsumoto2021detection}. The innovative design improves the capacity of the system to capture acoustic signals more accurately, as evidenced by measurements from an anechoic tank experiment.}

\item[(ii)] {We introduce the broadband generalized sparse covariance fitting approach, which effectively adapts to DAS arrays. This approach improves robustness in target detection and localization, even in the presence of non-uniform noise, with the simulation results confirmed. The term "robust" in the title specifically denotes resilience against spatially non-uniform environmental noise.}

\item[(iii)] {The off-grid challenge is tackled by integrating a grid refinement strategy that iteratively updates the steering dictionary, thereby enhancing the resolution of DOA estimates.}

\item[(iv)] {We deployed the DAS system in a lake for underwater acoustic monitoring. Analysis of the experimental data confirms that the broadband $q$-SPICE schemes can effectively enhance the detection and tracking capabilities of non-cooperative targets.}
\end{itemize}

The remainder of this paper is organized as follows: Section \ref{Sec2} introduces our DAS system and the signal model for underwater DAS arrays. Section \ref{Sec3} details the proposed broadband DOA estimation framework, with particular emphasis on its robustness to non-uniform noise environments. Section \ref{Sec4} analyzes the simulation results, validating the algorithm's performance under various operational conditions. Section \ref{Sec5} demonstrates the experimental validation through comprehensive tank and lake trials using field-collected data. Finally, section \ref{Sec6} concludes the paper.

\section{Background} \label{Sec2}
\subsection{DAS System discription}
The DAS technology utilizes the principle of coherent Rayleigh scattering and employs $\varphi$-OTDR for acoustic signal detection. In summarizing the basic principles of the $\varphi$-OTDR, we present it as follows: The coherent light pulses from a narrow-linewidth laser (optical source) are launched into the sensing fiber. The Rayleigh backscattering caused by microscopic variations in the fiber scatters the light back to the receiver (the end for detecting backscattering light). External acoustic vibrations or perturbations applied to the fiber optic cable result in its extension or contraction, causing a variation in the optical phase of the backscattered light. By monitoring these phase changes in the backscattered signal, we can realize highly sensitive distributed acoustic sensing with spatial resolution defined by the pulse width and sampling parameters. Regarding the workflow of our DAS system, we outline it briefly here as follows: 
\begin{itemize}
	\item Coherent detection is performed on the backscattered signal to retrieve its complex field representation, encompassing both amplitude and phase.
	\item The resulting phase data then undergoes an unwrapping algorithm to reconstruct its continuous variations along the fiber.
	\item Noise filtering is applied to enhance the signal quality and improve detection sensitivity.
	\item The time-of-flight differences of the optical pulses are mapped onto spatial coordinates to localize the acoustic events.
	\item Finally, the processed phase signals are analyzed in both the time and frequency domains to characterize and classify the external sounds or vibrations.
\end{itemize}

\begin{figure*}[!t]
	\centerline{\includegraphics[width=0.85\textwidth]{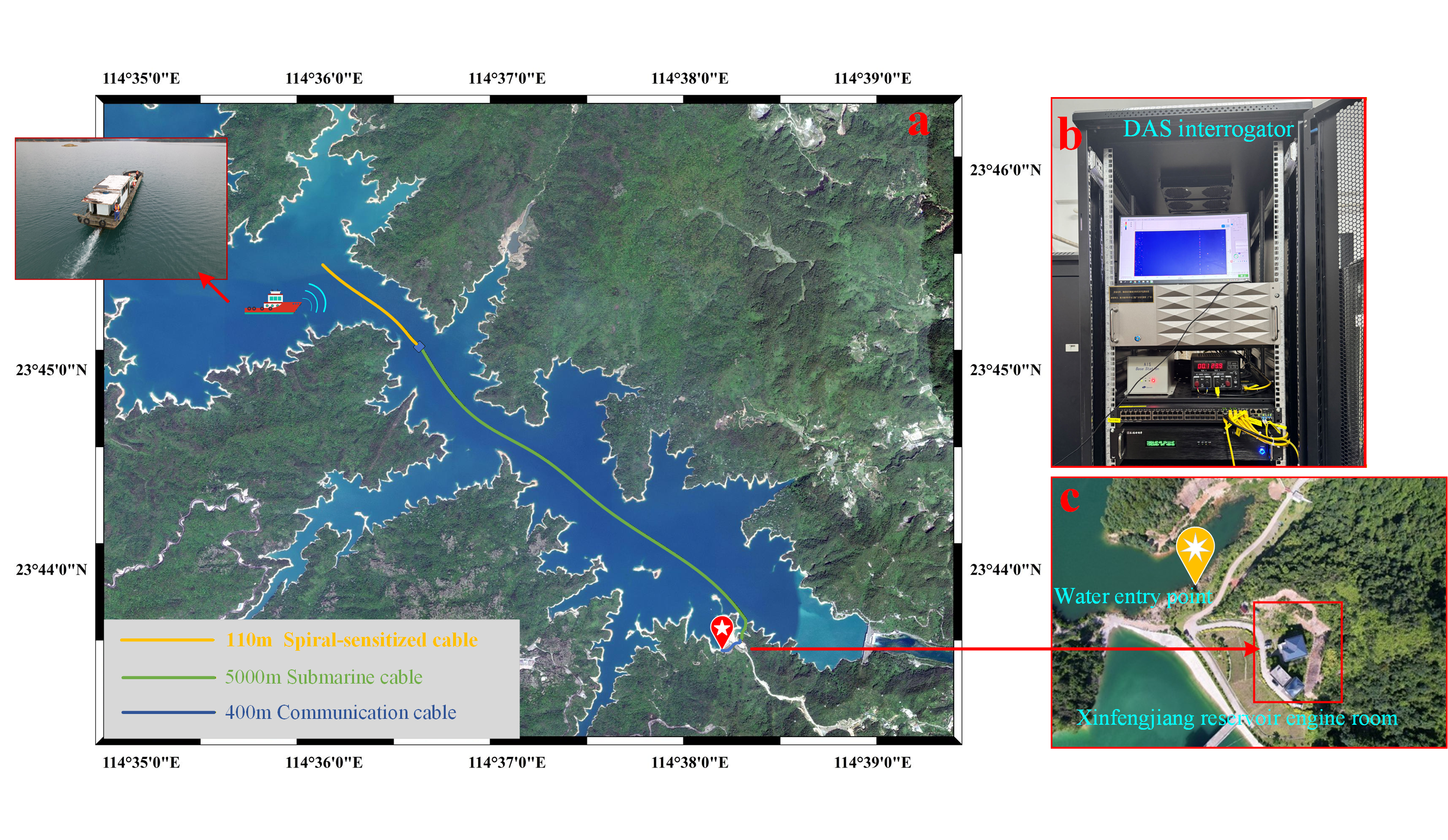}}
	\caption{Deployment diagram of the DAS system layout. (a) Satellite image of the Xinfengjiang Reservoir, showing the DAS optical cable route. (b) DAS interrogator. (c) Satellite image of the Xinfengjiang Reservoir engine room and the water entry point of the optical cable.}
	\label{Fig:DAS_deploy}
	\vspace{-3mm}
\end{figure*}
\begin{figure}[t]
	\centering
	\includegraphics[width=\columnwidth]{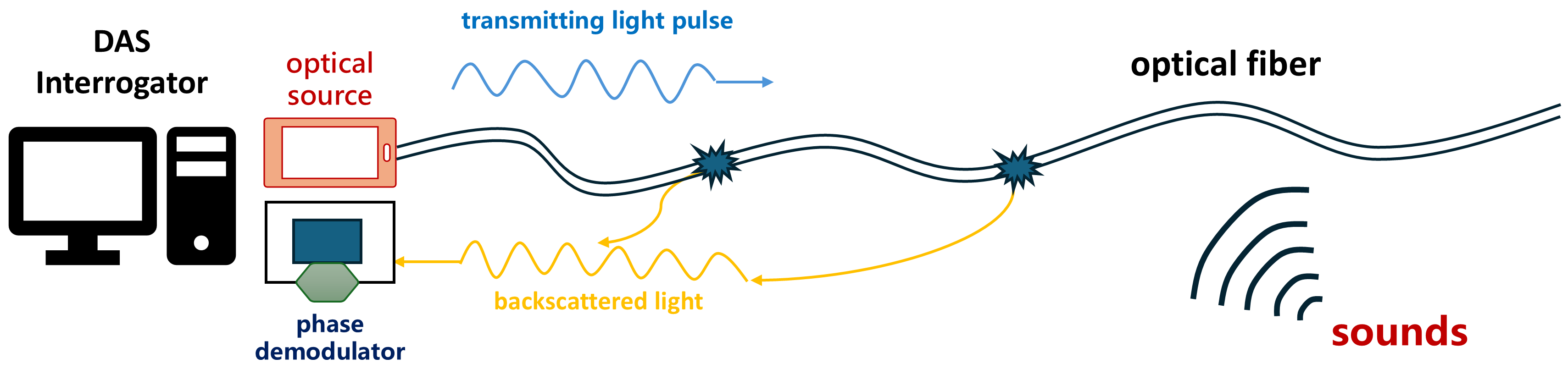}
	\caption{The schematic diagram of the $\varphi$-OTDR.}
	\label{Fig:OTDR}
	\vspace{-3mm}
\end{figure}

\begin{figure}[t]
	\centering
	\includegraphics[width=\columnwidth]{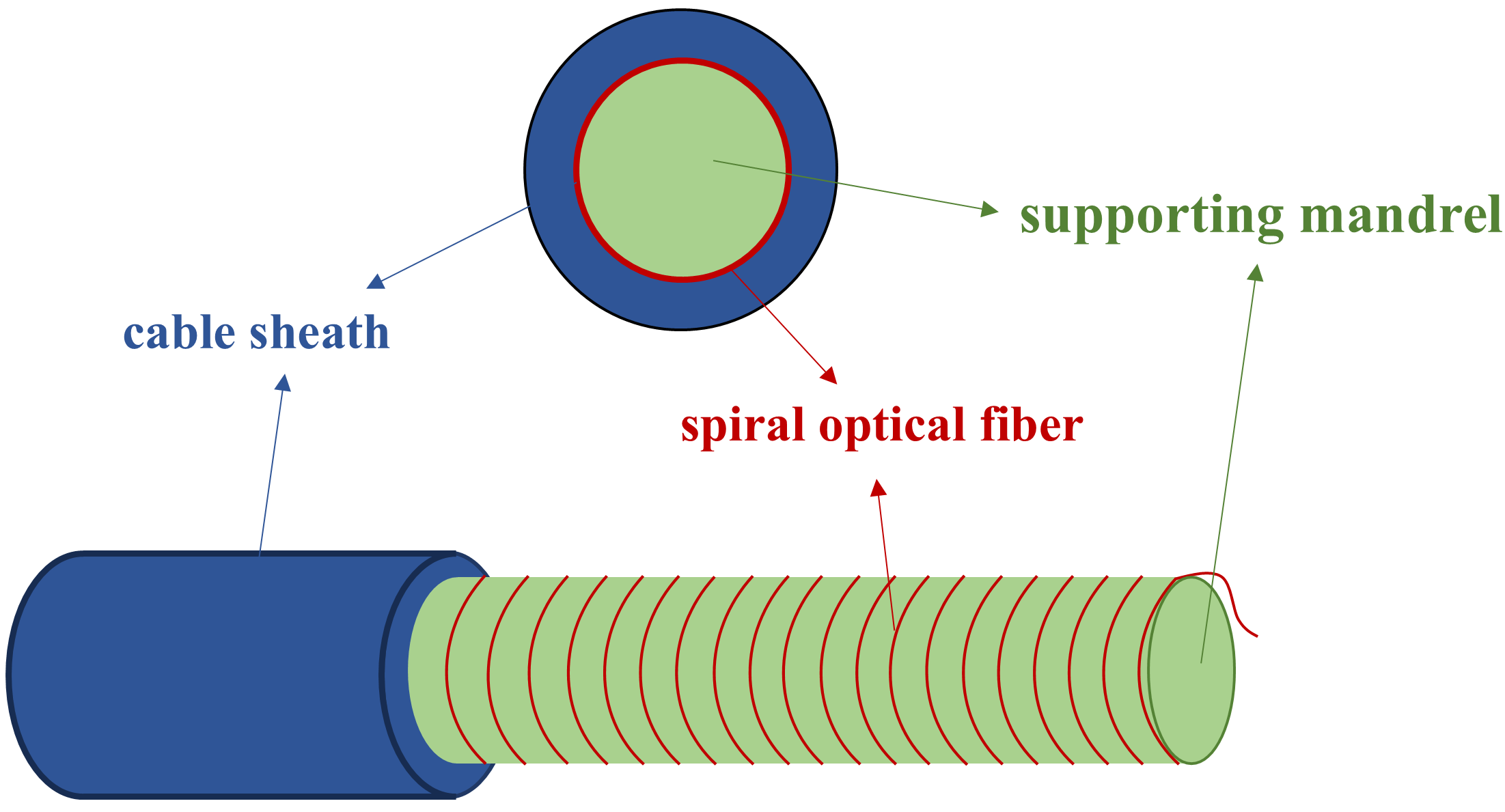}
	\caption{The schematic diagram of the spiral-sensitized optical cable.}
	\label{Fig:SpCable}
	\vspace{-3mm}
\end{figure}

In July 2022, our research team deployed an integrated land-water DAS system in the Xinfengjiang Reservoir, Heyuan City, Guangdong Province. The system was upgraded sequentially in May 2023, involving the attachment of a spiral-sensitized optical cable to the tail end of the original submarine cable. The deployment diagram of the DAS system layout is shown in Fig.\ref{Fig:DAS_deploy}. The schematic diagram of the $\varphi$-OTDR's operational principle is depicted in Fig.\ref{Fig:OTDR}. This enhancement aimed to explore and validate the feasibility of DAS technology for underwater acoustic monitoring, particularly in detecting weak acoustic signals.

The DAS interrogator was installed in the equipment room of the Xinfengjiang Management Office. The connected fiber-optic system consists of three segments: a \SI{400}{\meter} communication optical cable, a \SI{5}{\kilo\meter} standard armored submarine cable, and a \SI{110}{\meter} spiral-sensitized optical cable. In Fig.\ref{Fig:SpCable}, a schematic of the spiral-sensitized optical cable is depicted, demonstrating its spiral-wound structure for the optical fiber. Notably, the spiral-sensitized optical cable refers to the uniform circular winding of fiber around the mandrel.

\begin{figure}[t]
	\centering
	\includegraphics[width=0.45\textwidth]{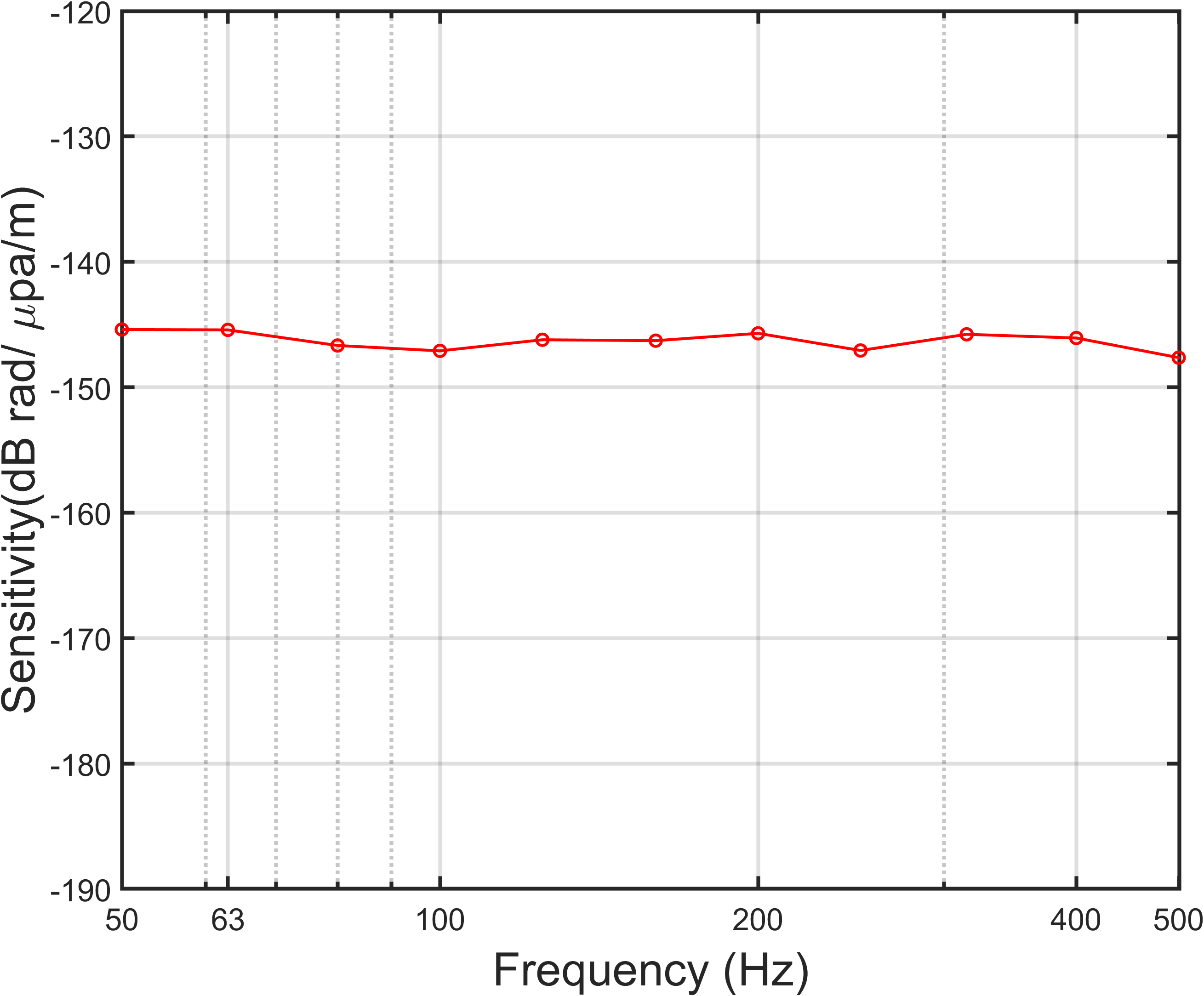}
	\caption{The sensitivity of the spiral-sensitized optical cable inside the standing-wave tube.}
	\label{Fig:sensitivity}
	\vspace{-3mm}
\end{figure}

This novel sensitized optical cable comprises a supporting mandrel, special bending resistant fiber and cable sheath. The supporting mandrel is made of acoustically sensitive material, tightly wound with the bend-resistant optical fibers. Due to the significantly lower Young's modulus of the support mandrel material compared to that of the optical fibers, this method greatly enhances the sound pressure sensitivity. The final diameter of the optical cable is approximately \SI{20.3}{\milli\meter}, with an average optical fiber length of about \SI{6.3}{\meter} wound per meter of the cable. Given a fiber length of $L$ wound around a cable of length $L_c$, the phase change of the round-trip transmission light induced by variations in the fiber length can be denoted as $\Delta\varphi=2 \varsigma \cdot  \Delta L$. And the acoustic pressure sensitivity of the cable can be obtained as follows \cite{jiajing2019distributed, lu2021distributed}:
\begin{align}
   M_S &=20 \mathrm{log}_{10}\left(\frac{\Delta\varphi}{P\cdot L_{c}}\right)  \\
          &=20 \mathrm{log}_{10}\left(\frac{2 \varsigma }{ P\cdot L_{c}}\Delta L\right) 
\end{align} where $\varsigma$ is denoted as the propagation constant, then it can be formulated by $ \varsigma =\frac{2 \pi n}{\lambda_L} $, then we get
\begin{align}
	M_S &=20 \mathrm{log}_{10}  \left(  \frac{4\pi n}{\lambda_L P\cdot L_{c}}  \Delta L   \right) \\
	       &=20 \mathrm{log}_{10} \left(\frac{4\pi n}{\lambda_L P\cdot L_{c}}\frac{\Delta r}{r}L\right)
\end{align}
Here, $\Delta L$ represents the change in the length of the optical fiber.  $\lambda_L$ signifies the wavelength of light. $n$ represents the refractive index. $P$ is the sound pressure on the cable, $r$ is the radius and $\Delta r$ is the radial displacement of the supporting mandrel. Through mechanical material analysis, the radial displacement of the used mandrel under specific internal and external pressures can be mathematically described as 
\begin{align}
	\Delta r=\frac{1-\varrho}{V}\cdot\frac{a^2p_1-r^2p_2}{r^2-a^2}\cdot r+\frac{1+\varrho}{V}\cdot\frac{a^2r^2(p_1-p_2)}{r^2-a^2}\cdot\frac{1}{r}
\end{align}
Where $\varrho$ and $V$ represent the Young's modulus and Poisson's ratio of the mandrel, while $a$ and $r$ denote the internal and external radius of the mandrel, and $p_1$ and $p_2$ stand for the internal and external pressure. Additionally, the sensitivity of the sensitized optical cable was evaluated within the frequency range of 50–500 Hz. The cable was wound into a loop and positioned at the same location as the piezoelectric (PZT) hydrophone inside the standing-wave tube. Utilizing the PZT hydrophone, which is the calibrated standard hydrophone, as the reference for sensitized optical cable calibration. Based on the phase change demodulated by the $\varphi$-OTDR system and the corresponding sound pressure measured by the PZT hydrophone, the sensitivity of the sensitized optical cable was determined. The experimental results are presented in Fig. \ref{Fig:sensitivity}. Hence, the acoustic pressure sensitivity is approximately -145.69 dB $ \emph{re:} \si{1\radian / (\micro\pascal \cdot \meter)} $.

In addition to the optical interrogation mechanism, the sensitivity of the spiral-wound sensitized cable is strongly influenced by its composite structure. Critical factors include the optical fiber type, mandrel material and configuration (solid or air-backed), and the presence of a polyurethane (PU) coating or molding layer, all of which affect acoustic coupling efficiency. These structural elements not only alter the effective refractive index distribution but also govern the sensor's mechanical response. The latter is determined by the elastic properties of the composite materials, such as Young's modulus and Poisson's ratio, which control the transfer of external acoustic pressure to the embedded optical fiber. The material selected for the support mandrel in this study has a Young's modulus of 1.0 GPa and a Poisson's ratio of 0.4. Consequently, the sensitivity reported encompasses both the optical and mechanical effects of this particular configuration.
The total cable length is approximately \SI{5.51}{\kilo\meter}. This entire cable was transformed into a sensor array comprising 620 virtual acoustic sensing channels. The average spacing between sensing points is \SI{10}{\meter} for the segment composed of the communication fiber and standard submarine cable, while it is reduced to approximately \SI{1.375}{\meter} for the spiral-sensitized segment, allowing for enhanced sensitivity in that section. The spatial resolution is influenced not only by the system settings (e.g., pulse width) but also by the gauge length. The actual spatial resolution should indeed be the smaller of the two factors.

The cross-sectional profile of the submarine cable layout is illustrated in Fig.\ref{Fig:cros_Fig}. The cable enters the water at approximately \SI{500}{\meter} from the DAS interrogator and then gradually sinks to the bottom of the reservoir due to its negative buoyancy. It eventually couples with the sediment at the reservoir floor, ensuring acoustic coupling. The technical specifications of the DAS interrogator are summarized in TABLE~\ref{tab:Table00}. 

\begin{table}[!t]
	\caption{The specifications of the DAS system, including the interrogator and three types of sensing cables.\label{tab:Table00}}
	\centering
	\begin{tabular}{|c|c|} 
		\hline 
		Indicators & Parameters\\
		\hline  
		Weight&10 kg\\
		
		Volume&60 cm × 30 cm × 20 cm\\
		
		Maximum length of monitoring& $\geq$ 70 km\\
		
		Measuring frequency band& 0.01 Hz $\sim$ 2 kHz\\
		
		Sampling frequency& $\geq$ 5 kHz\\
		
		Spatial resolution&\SI{1}{\meter} $\sim$ \SI{10}{\meter} (Optional)\\
		
		Operating temperature& $-10^\circ\mathrm{C} \sim 55^\circ\mathrm{C}$\\

        Power consumption & $\leq$ 240 $W$ \\
        
        Gauge length & \SI{10}{\meter} \\
		\hline
	    Communication optical cable & Parameter\\
	    \hline 
	     Sensitivity& -202 dB $ \emph{re:} \si{1\radian / (\micro\pascal \cdot \meter)} $\\
	     \hline
	     Standard submarine cable & Parameter\\
	     \hline 
	     Sensitivity& -202 dB $ \emph{re:} \si{1\radian / (\micro\pascal \cdot \meter)} $\\
	    \hline 
	    Spiral sensitized optical cable & Parameter\\
	    \hline 
	    Sensitivity& -145.69 dB $ \emph{re:} \si{1\radian / (\micro\pascal \cdot \meter)} $\\
	    \hline 
	\end{tabular}
\end{table}

\begin{figure*}[!t]
	\centerline{\includegraphics[width=0.8\textwidth]{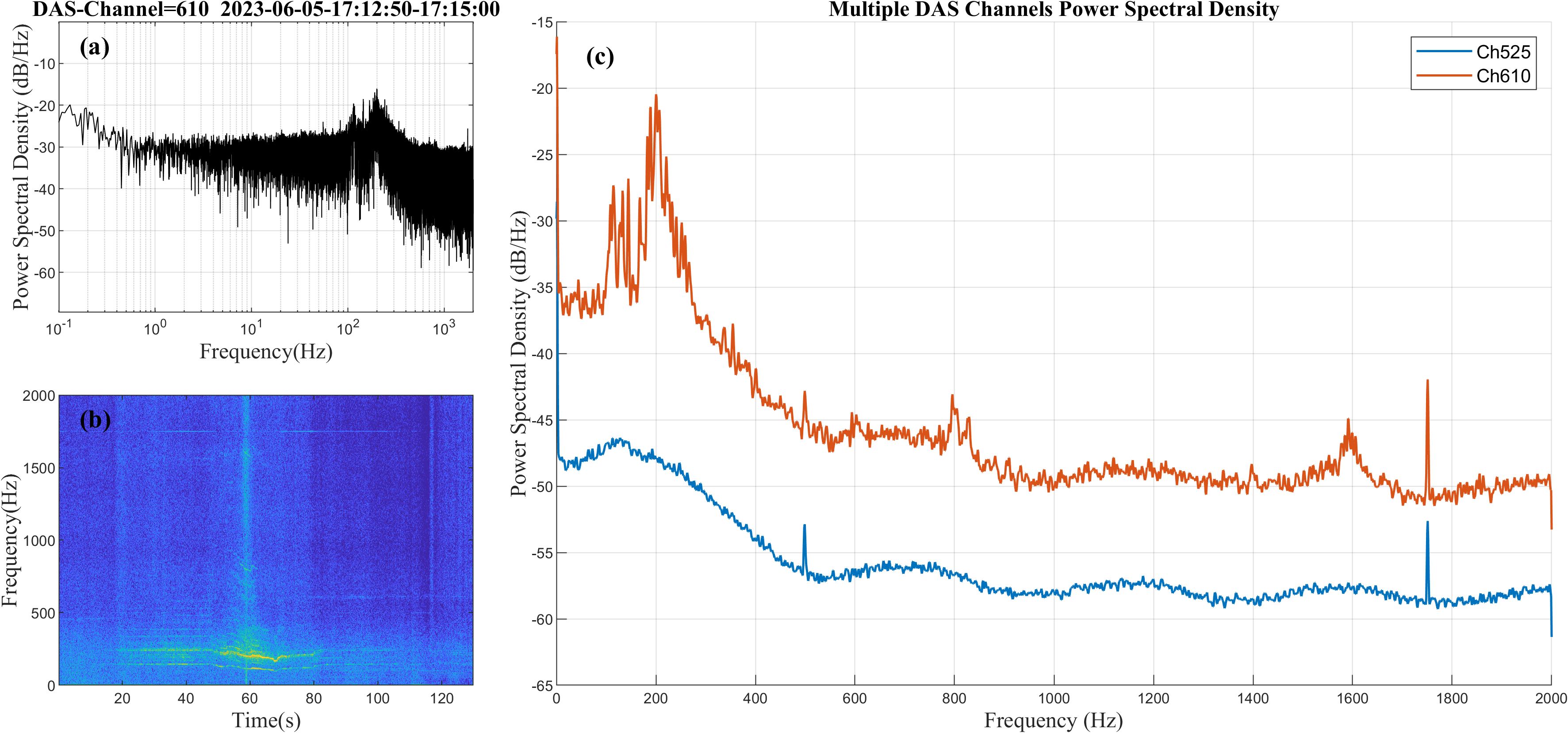}}
	\caption{Analysis of acoustic signals from conventional and high-sensitivity DAS cables. (a) Power spectral density (PSD) of the signal from Channel 610 (high-sensitivity cable). (b) Time-frequency analysis of the signal from Channel 610, showing the radiated noise of a vessel's propeller. (c) Comparison of the PSD between Channel 525 (conventional cable) and Channel 610.}
	\label{Fig:SignalGainCompare}
	\vspace{-3mm}
\end{figure*}

To quantitatively compare the signal gain between conventional and high-sensitivity cables, data from June 5, 2023, were selected for analysis. This period includes signals recorded when a vessel operating nearby started its engine. The power spectral density analysis was performed on these signals. As shown in Fig.\ref{Fig:SignalGainCompare}, Channel 525 corresponds to the conventional cable, while Channel 610 corresponds to the high-sensitivity cable. Fig.\ref{Fig:SignalGainCompare}a and \ref{Fig:SignalGainCompare}b present the power spectrum and time-frequency analysis results for Channel 610, clearly revealing the radiated noise from the vessel's propeller. Fig.\ref{Fig:SignalGainCompare}c also indicates that the high-sensitivity cable provides an additional 15 dB of signal gain compared to the conventional cable.

\subsection{Signal Model}
Consider an $M$-element horizontal linear array of underwater DAS sensors passively measuring the radiation signals from $K$ far-field narrowband sources, $ \{ s_k(t) \}$, impinging from directions  $\{ \theta_k \}$, 
for $k=1,\dots, K$. Assuming the source signals to be statistically uncorrelated, the received signal may be expressed as \cite{liu2022robustVBI}
\begin{align}
	\textbf{y}_t & =\sum_{k=1}^{K}\textbf{a}\left(\theta_{k}\right)s_{k}\left(t\right)+\textbf{e}_t \nonumber \\ 
	&=\textbf{A}_\theta\textbf{s}_t+\textbf{e}_t \quad  \in \mathbb{C}^{M\times 1}
	\label{eq:yt}
\end{align}
for $t=1,\ldots, N$, where the array manifold and the source signals at time $t$ are defined as
\begin{align}
	\textbf{A}_\theta &= \left[ \begin{array}{ccc}  \mathbf{a}(\theta_1) & \dots & \mathbf{a}(\theta_K)\end{array} \right] \in \mathbb{C}^{M\times K} \\
	{\bf s}_t &= \left[ \begin{array}{ccc} s_1(t) & \ldots & s_K(t) \end{array} \right]^T \in \mathbb{C}^{K\times 1}
\end{align}
respectively, with ${\bf e}_t \in \mathbb{C}^{M\times 1}$ denoting the additive noise vector at time $t$. Here, the steering vector, $\textbf{a}\left(\theta_{k}\right)$, in the case of perfectly uniform array, takes the form 
\begin{align}
	\mathbf{a}(\theta_k) &= \left[ \begin{array}{cccc} 1 &e^{\frac{-j d \omega_c \sin(\theta_k)}{c}} & \ldots & e^{\frac{-j d (M-1) \omega_c  \sin(\theta_k) }{c}} \end{array} \right]^T \in \mathbb{C}^{M\times 1}
	 \label{eq:steer}
\end{align}
where $\omega_c = 2\pi f_c$, with $f_c$ denoting the carrier frequency, $d$ the (nominal) sensor spacing, generally set to be half the sound wavelength, $\lambda$, and $c$ the speed of the sound propagation under water. As this propagation speed will depend locally on various aspects of the underwater environment, such as temperature, salination, currents, and depths, the sound speed may only be treated as approximately known, at best. 
In this work, we assume that the source signals, ${\bf s}_t$, and the additive noise, ${\bf e}_t$, are uncorrelated, with the noise being well modelled as being zero-mean (circularly symmetric) Gaussian distributed. 

\section{Proposed Method} \label{Sec3}
\subsection{The array $q$-SPICE estimator} 
It may be noted that the covariance matrix of ${\bf y}_t$ in \eqref{eq:yt} may be expressed as \cite{stoica2010spice}
\begin{align}
	{\bf R}_{\tilde{\bf p}} &= E\left\{\textbf{y}_t \textbf{y}^\text{H}_t \right\}
	= \boldsymbol{A}_\theta \boldsymbol{P_s} \boldsymbol{A}_\theta ^\text{H}+\boldsymbol{\Sigma}_e \nonumber \\
	&= \boldsymbol{A}_\theta\left[\begin{array}{ccc}
		\sigma_{s_1}^2 &  &  \\
		& \ddots &  \\
		&  &  \sigma_{s_K}^2  \\
	\end{array}\right]\boldsymbol{A}_\theta^\text{H}+\left[\begin{array}{ccc}
		\sigma_{e_1}^2 &  &  \\
		& \ddots &   \\
		&  & \sigma_{e_M}^2  \\
	\end{array}\right] \nonumber \\
	&=\boldsymbol{B} \boldsymbol{\Sigma}_{\Theta} \boldsymbol{B}^H \label{eq:Rp}
\end{align}
where $E\{ \cdot \}$ denotes the statistical expectation, $(\cdot)^H$ the conjugate transpose, $\sigma_{s_k}^2$ and $\sigma_{e_m}^2$ the power of the $k$th signal and the $m$th noise component, respectively, and
\begin{align}
	\boldsymbol{B} &=[\boldsymbol{A}_\theta \quad \boldsymbol{I}_M ] \\
	\boldsymbol{\Sigma}_{\Theta} & =\text{diag}\Bigl\{ \left[ \begin{array}{cccccc} \sigma_{s_1}^2 & \dots & \sigma_{s_K}^2 & \sigma_{e_1}^2 & \ldots &\sigma_{e_M}^2 \end{array} \right] \Bigr\} \nonumber \\
	& =\text{diag}\Bigl\{ \left[ \begin{array}{cc} {\bf p}^T & \boldsymbol{\sigma}^T \end{array} \right] \Bigr\} \label{eq:Sigma_Th}
\end{align}
with $\text{diag}\{ \cdot \}$ denoting the diagonal matrix formed with the given vector along its diagonal, ${\bf I}_M$ the $M \times M$ identity matrix, 
and
\begin{align}
	{\bf p} &= \left[ \begin{array}{ccc} \sigma_{s_1}^2 & \dots & \sigma_{s_K}^2 \end{array} \right]^T \\
	\boldsymbol{\sigma} &= \left[ \begin{array}{ccc}\sigma_{e_1}^2 & \ldots &\sigma_{e_M}^2  \end{array} \right]^T   \\
	\tilde{\bf p} &= \left[ \begin{array}{cc} {\bf p}^T & \boldsymbol{\sigma}^T  \end{array} \right]^T   \label{eq:tildep}
\end{align}
In a marine environments, the ambient background noise is generally non-uniform, such that each sensor typically experience noise variance, such that 
\begin{equation}
	\sigma_{e_1}^2 \neq \sigma_{e_2}^2 \neq \dots \neq \sigma_{e_M}^2	
\end{equation}

\subsection{Allowing for broadband signals} 
In an underwater environment, it is often difficult to maintain a fully calibrated uniform sensing array, especially for sensors that are not physically attached to each other. Furthermore, because of the deployment process and the irregular seafloor terrain, even for cases when the sensing array is physically connected in a uniform manner, the sensors are rarely perfectly calibrated or precisely spaced at the assumed half-length interval. For these reasons, we now proceed to assume the $M$ sensors are places in a reasonably linear manner, but allow their positioning to deviate somewhat from a uniform linear array configuration. Thus, we extend the earlier model to incorporate potential errors in the positioning of the array elements, such that
\begin{align}
	\tilde{{\bf a}}(\theta_k) &= \left[ \begin{array}{cccc} 1 &e^{-j \alpha_1 d \omega_c \sin(\theta_k) / c} & \ldots & e^{-j \alpha_{M-1} d \omega_c  \sin(\theta_k) / c} \end{array} \right]^T 
\end{align}
where $\alpha_{m-1} d$ represent the distance between the $m$th sensing element and the (first) reference element. As shown in \eqref{eq:steer}, $\alpha_{m-1}$ would thus take integer values for an ULA, which is here relaxed to allow for arbitrary positive real values, i.e., $\alpha_{m-1} \in \mathbb{R}$.

It should further be noted that the earlier presentation is restricted to narrowband sources. Many forms of sonar measurements are predominantly broadband, such as those resulting from ships, which typically consists of 
a broad continuous spectrum overlayed by multiple spectral lines, or due to the frequency dispersion effects in the underwater acoustic channel that naturally widens narrowband sources. 

In such cases, the receiver typically applies a bandpass filterbank separating the source contribution into a set of narrowband signals. Consider a set of frequency grid points 
\begin{align}
	\left\{ \frac{2 \pi \ell_1 }{ N}, \frac{2 \pi \ell_2 }{ N}, \ldots, \frac{2 \pi \ell_P }{ N} \right\} 
\end{align}
with $\ell_1, \ldots, \ell_P$ being $P$ given, not necessarily consecutive, integers selected to cover the frequency range of interest. 
To express the measured signal over this selected frequency range, let
\begin{align}
	{\bf Y} &= \left[ \begin{array}{ccc} {\bf y}_1 & \ldots & {\bf y}_{N} \end{array} \right] 
	= \textbf{A}_\theta \textbf{S} + \textbf{E} \in \mathbb{C}^{M\times N}
\end{align}
where $\bf S$ and $\bf E$ are formed similarly to $\bf Y$. In practical array processing, it is common to assume $N \geq M$ and typically $N \geq M$, so that the sample covariance matrix can be estimated reliably and the separation between the signal and noise subspaces is statistically stable.

Let ${\bf Y}_{m:}(t)$ denote the $m$th row vector of $\bf Y$, which thus gathers the time-domain signal measured by the $m$th sensor, transforming each such vector to form
\begin{align}
	Z_{m,\ell} &=  {\bf Y}_{m:}^T(t) {\bf v}_\ell    
\end{align}
for $\ell=1,\ldots,P$, 
where
\begin{align}
	{\bf v}_\ell &= \left[ \begin{array}{cccc} 1 & z_\ell & \ldots & z_\ell^{{N}-1} \end{array} \right]^T 
\end{align}
with $z_\ell = e^{i 2 \pi \ell / {N}}$. The sequence $Z_{m,\ell}$ is thus the frequency content for the $m$th sensor at frequency index $\ell$. Notably, we explore an adaptive optimal frequency selection strategy based on eigenvalue discrimination for broadband DOA estimation of underwater moving targets with time-varying line spectra. By maximizing the separation between the signal and noise subspaces, our method automatically selects the most reliable frequency point in each processing frame to track these spectral variations. This adaptive mechanism operates without relying on a fixed frequency set. This design effectively mitigates Doppler-induced spectral mismatch and improves robustness under low SNR and limited snapshot conditions.

Combining the response for each sensor, 
\begin{align}
	{\bf Z}_\ell &= \left[ \begin{array}{c} Z_{1,\ell} \\\vdots \\ Z_{M,\ell} \end{array}\right] = \left[ \begin{array}{c}  {\bf Y}_{1:}^T(t) {\bf v}_\ell \\\vdots \\ {\bf Y}_{M:}^T(t) {\bf v}_\ell \end{array}\right] = \textbf{A}_\theta \textbf{S} {\bf v}_\ell + \textbf{E} {\bf v}_\ell 
\end{align}
such that the frequency selective representation of \eqref{eq:yt}
is
\begin{align}
	{\bf Z}_\ell &= \textbf{A}_\theta \tilde{\bf S}_\ell+ \tilde{\bf E}_\ell \in \mathbb{C}^{M\times 1}
\end{align}
for $\ell=1,\ldots,P$, where $\tilde{\bf S}_\ell = \textbf{S} {\bf v}_\ell$ and 
$\tilde{\bf E}_\ell = \textbf{E} {\bf v}_\ell$. Forming the covariance matrix of ${\bf Z}_\ell$ reminiscent to \eqref{eq:Rp} yields
\begin{align}
	{\bf R}_{\tilde{\bf p}_{Z}} &= E\left\{ {\bf Z}_\ell {\bf Z}_\ell^H \right\}
	= \boldsymbol{A}_\theta \boldsymbol{P}_{\bf Z} \boldsymbol{A}_\theta^H + \boldsymbol{\Sigma}_E \nonumber 
	=\boldsymbol{B} \boldsymbol{\Sigma}_{\Theta}^Z \boldsymbol{B}^H \label{eq:Rp2}
\end{align}
where, using definitions reminiscent of \eqref{eq:Sigma_Th}-\eqref{eq:tildep},
\begin{align}
	\boldsymbol{\Sigma}_{\Theta}^Z & =\text{diag}\Bigl\{ \left[ \begin{array}{cccccc} \sigma_{\tilde{s}_1}^2 & \dots & \sigma_{\tilde{s}_K}^2 & \sigma_{E_1}^2 & \ldots &\sigma_{E_M}^2 \end{array} \right] \Bigr\} \nonumber \\
	& =\text{diag}\Bigl\{ \left[ \begin{array}{cc} {\bf p}^T_{\tilde{s}} & \boldsymbol{\sigma}^T_E \end{array} \right] \Bigr\} 
	=\text{diag}\Bigl\{ \left[ \begin{array}{c} \tilde{\bf p}^T_{Z}  \end{array} \right] \Bigr\} 
\end{align}
One may then formulate the broadband $q$-SPICE estimator as
\begin{equation}
	\min_{{\boldsymbol{p_{\tilde s}}}\ge 0,\boldsymbol{\sigma}_E \ge 0} \mathbf{y}^\mathrm{H} 
	{\bf R}^{-1}_{\tilde{\bf p}_{Z}} \mathbf{y}+\left\|\boldsymbol{W}_Z \boldsymbol{p}_{\tilde s} \right\|_r + \left\|\boldsymbol{W}_{\sigma_E} \boldsymbol{\sigma}_E \right\|_q \\  \label{Eq23b}
\end{equation}
with the weight matrices formed correspondingly. 

\begin{figure}[!t]
	\centering
	\includegraphics[width=6cm, height=7cm]{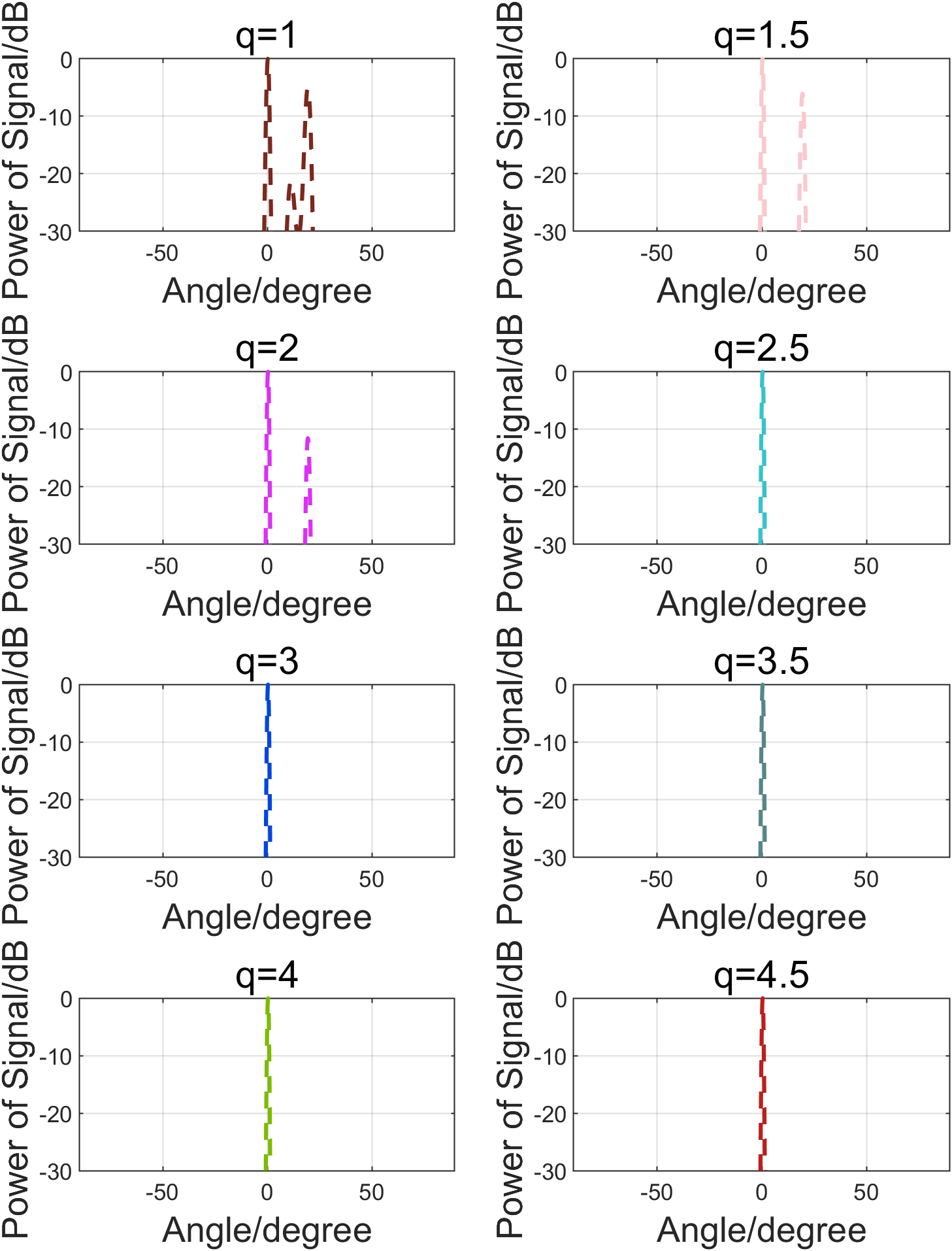}
	\caption{Spatial spectral estimates under different values of the parameter $q$.}
	\label{Fig:bestq}
	\vspace{-3mm}
\end{figure}

Evidently, one can acquire the original SPICE framework by setting $r=1, q=1$. Given that $r\geq 1$ and $q \geq 1$, the optimization problem stays convex. As given by (\ref{Eq23b}), the broadband $q$-SPICE algorithm separates the signal and noise terms by contrast with the original formulation. The exploitation of the $q$-norm penalty on the noise leads to a precise estimation of the noise variance, consequently ensuring that $\mathbf{R}$ attains full rank. Fig.\ref*{Fig:bestq} illustrates the spatial spectral estimates of the $q$-SPICE algorithm under different values of $q$. The results demonstrate that $q > 2$ induces missed detections, whereas $q = 1$ manifests an undesirably non-sparse spectral profile. Based on this trade-off, the operational range of $q \in [1.5, 2]$ is proposed to ensure robust and sparse source estimation.

\subsection{Refined Dictionary for off-grid DOAs}
\begin{figure}[t]
	\centering
	\includegraphics[width=\columnwidth]{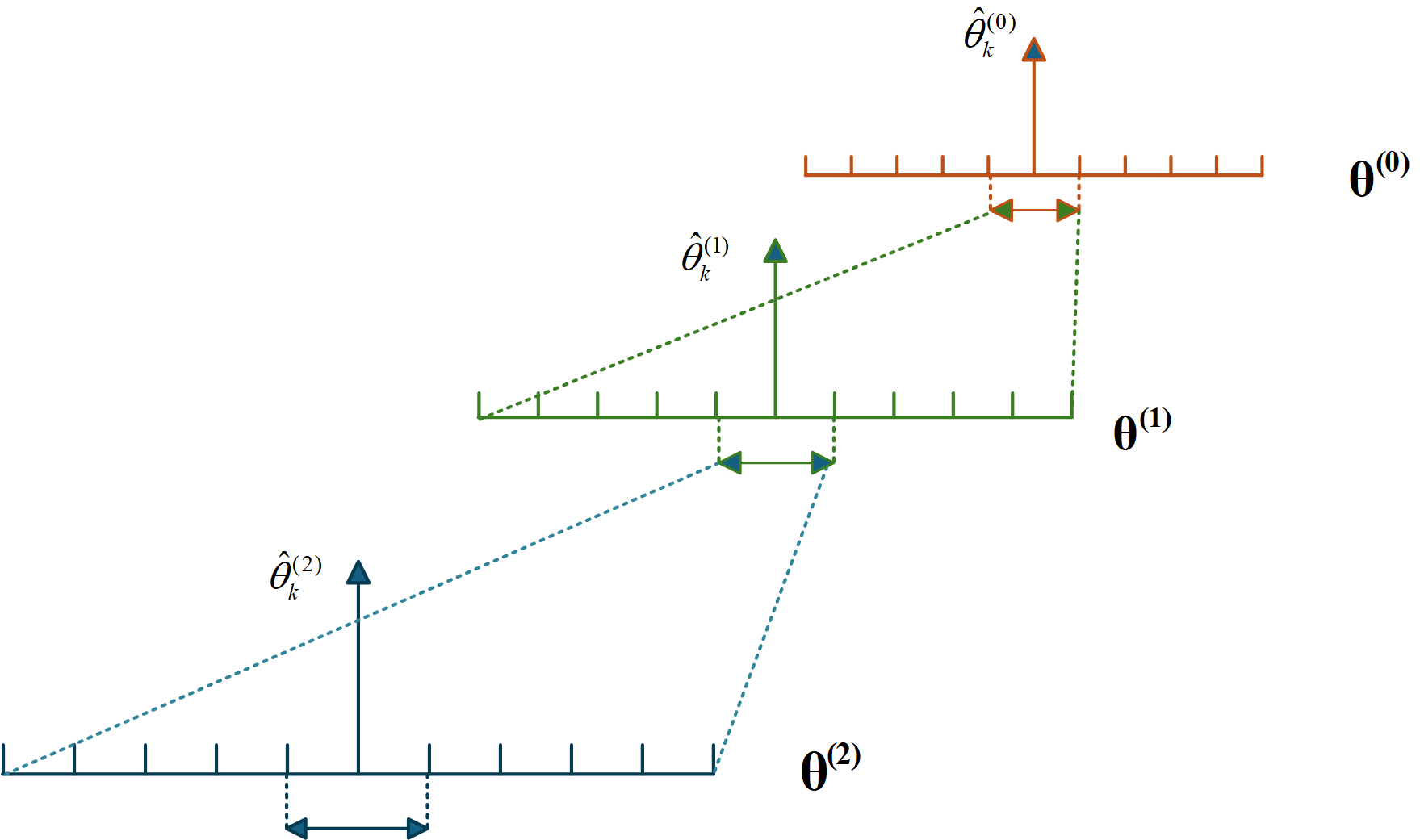}
	\caption{Schematic illustration of $\text{GNR}^2$.}
	\label{Fig:GNR}
	\vspace{-3mm}
\end{figure}

In the earlier discussion, it was assumed that the true direction is established on the grid of the dictionary. However, this is not always the case. The off-grid DOA estimating presents a significant research challenge across various applications. In this article, we explore a straightforward approach that employs a redefined dictionary updating technique to enhance the resolution of the estimates. We term this approach as Grid Neighborhood Refined and Re-estimate($\text{GNR}^2$), which initially proposed by Dmitry Malioutov~\cite{malioutov2005sparse}.
The basic principle of $\text{GNR}^2$ involves initially dividing the azimuth domain using a coarse grid. Once an initial DOA estimate is obtained, the neighborhood azimuth around the selected grid is further subdivided, thus generating an updated dictionary (new array manifold matrix) with a finer grid. Fig.\ref{Fig:GNR} illustrates the schematic of $\text{GNR}^2$. This iterative process, known as $q$-SPICE-$\text{GNR}^2$, enables improved estimation accuracy with reduced computational complexity, as described in the following
\begin{itemize}
	\item To begin with, the initial region of grid is established and set to be divided. When dealing with DAS arrays, it is common practice to divide the entire observation area using either a 0.5 degree or 1 degree interval. This process yields an on-grid dictionary, which is subsequently employed for DOA estimation through the $q$-SPICE method.
	\item Once the estimated $\hat{\theta}$ is obtained, we will divide the neighborhood of $\hat{\theta}$ on both sides and then proceed with refinement. After obtaining the new dictionary, the DOA estimation will be conducted again using the $q$-SPICE method.
	\item Repeat the aforementioned steps until the estimated accuracy meets the predetermined requirements.
\end{itemize}
\section{Numerical Example} \label{Sec4}
In this section, we proceed to evaluate the performance of the proposed underwater target remote sensing framework by comparing it with several target bearing sensing techniques. During the evaluation process, we show the performance of $q$-SPICE and $q$-SPICE-$\text{GNR}^2$ in comparison to conventional beamforming (CBF), MUSIC, sparse bayesian learning (SBL)~\cite{gerstoft2016multisnapshot}, and the original SPICE\cite{stoica2010spice}.
	
The signal-to-noise ratio (SNR) for uniform noise are defined as \cite{shao2019optimal}
\begin{equation}
	\text{SNR}=10\log_{10}\left(\frac{P_{y}}{\sigma^{2}}\right) \mathrm{dB}
\end{equation}
where $P_{y}$ denotes the power of the signal and $\sigma^{2}$ the variance of the additive noise. The SNR for non-uniform noise \cite{Guo2023Variational} is defined as 
\begin{equation}
	\text{SNR}_{nu}=10\log_{10}\left[\frac{P_{y}}M\sum_{m=1}^M\left(1/\sigma_m^2\right)\right]\mathrm{dB}
\end{equation}
where $\sigma_m$ denotes the noise variance at $m$-th element.

\subsection{Spatial power spectrum comparison}\label{section4.1}
In our initial numerical analysis, we evaluate how diverse algorithms perform in estimating spatial spectrum with a 12-element underwater sensor array under 0dB SNR in two scenarios. The first case involves uniformly Gaussian spatial noise and a uniform array with half-spacing wavelength. The second scenario includes non-uniform Gaussian spatial noise and arbitrary linear array configurations. In the case of non-uniform noise, the main diagonal element of the covariance matrix is precisely defined as $[12, 2.3, 20.5, 5.5, 11.1, 6.5, 2, 13.5, 0.8, 1.7, 13.6, 5.2]$. For the arbitrary linear array configuration, the array positions were specified as [0; 1.9; 3.5; 5.3; 7.85; 10; 11.8; 14.1; 16; 17.7; 20; 22]$*d$. Here, $d$ denotes half the wavelength of the transmitting signal. The two tonal frequencies is 3000 Hz and 3100 Hz.
\begin{figure*}[!t]
	\centerline{\includegraphics[width=0.75\textwidth]{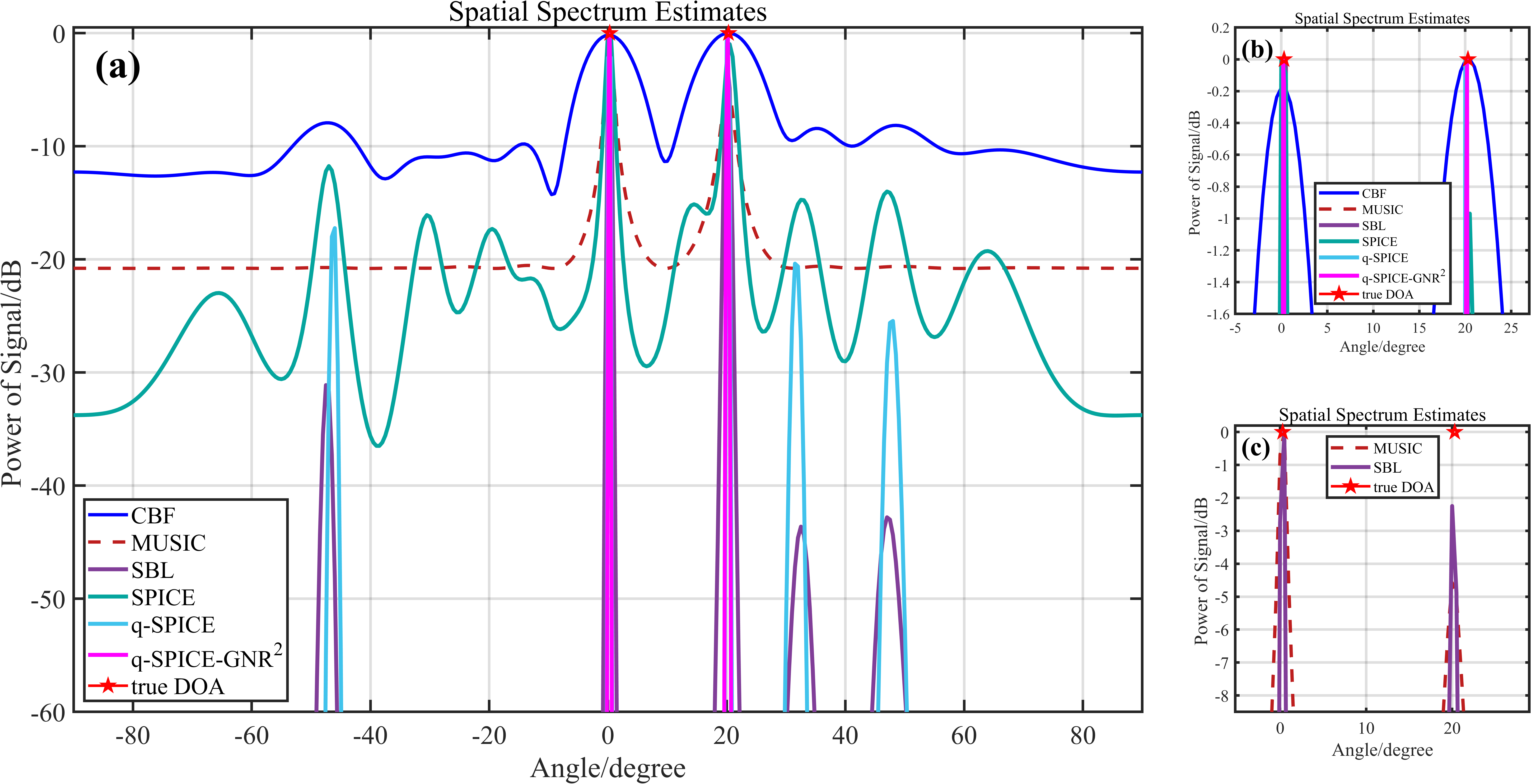}}
	\caption{The estimated spatial power spectrum for the case of uniformly Gaussian spatial noise, using uniform array with half-wavelength spacing. (a) The spatial power spectrum result provided by all approaches; (b) Zoom drawing with showing detail from $-1^{\circ}$ to $26^{\circ}$; (c) Zoom drawing for MUSIC and SBL algorithm.}
	\label{Fig:estDOA_unifNois}
	\vspace{-3mm}
\end{figure*}

\begin{figure*}[t]
	\centerline{\includegraphics[width=0.75\textwidth]{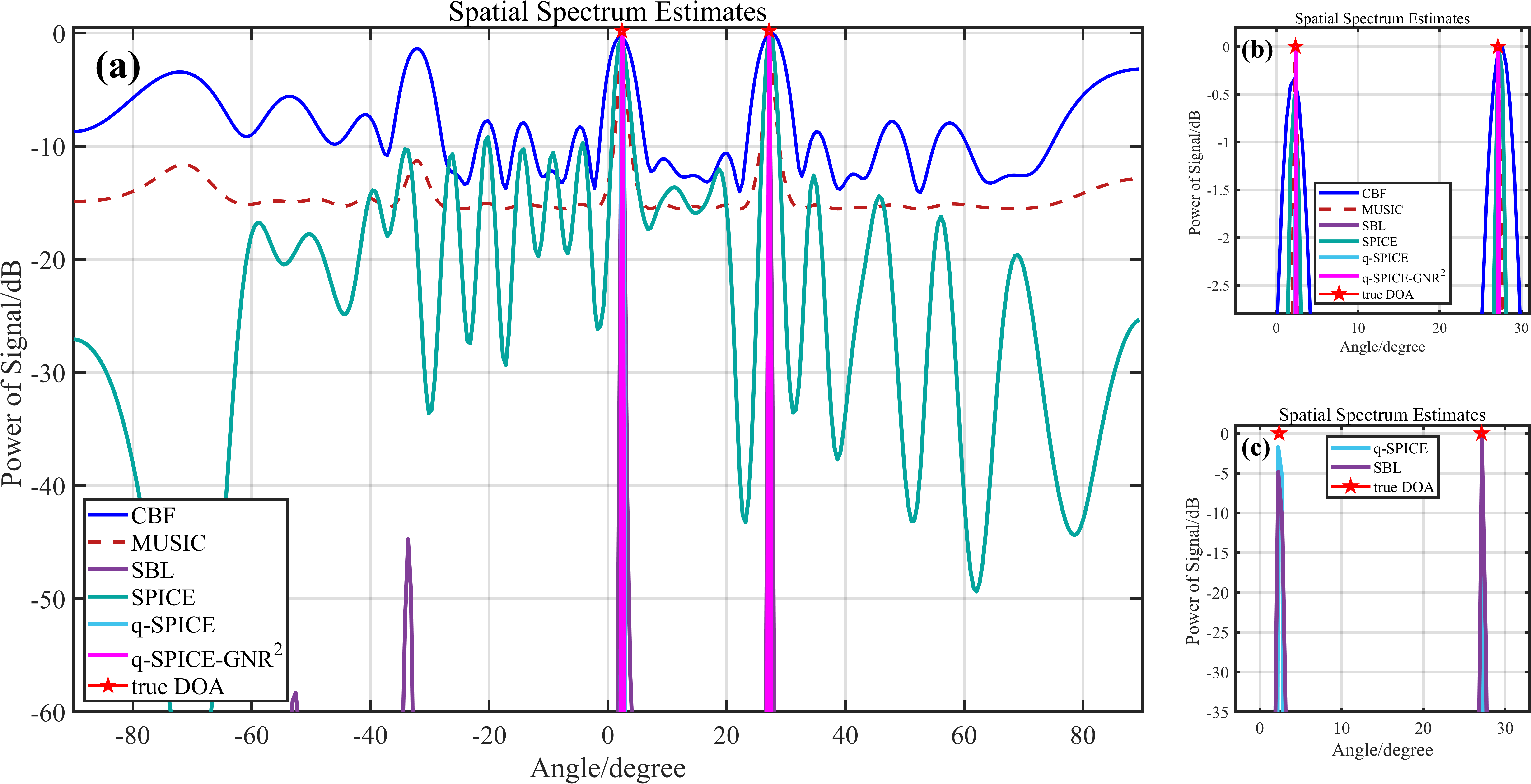}}
	\caption{The estimated spatial power spectrum for the case of non-uniformly Gaussian spatial noise, with array position errors. (a) The global spatial power spectrum result; (b) Zoom drawing with showing detail from $-1^{\circ}$ to $31^{\circ}$; (c) Zoom drawing for q-SPICE and SBL algorithm.}
	\label{Fig:estDOA_Non_unifNois}
	\vspace{-3mm}
\end{figure*}

Fig.\ref{Fig:estDOA_unifNois}a shows the estimated spatial power spectrum obtained from all methods for uniformly Gaussian spatial noise, utilizing a uniform array with half-wavelength spacing. Fig.\ref{Fig:estDOA_unifNois}b illustrates a zooming view of the azimuth region spanning from $-5.5^{\circ}$ to $26^{\circ}$. To enhance visibility of the spatial spectrum details obtained through the MUSIC and SBL methods, Fig.\ref{Fig:estDOA_unifNois}c presents individual results for these two approach. As illustrated in the Fig.\ref{Fig:estDOA_unifNois}, all techniques are capable of sensing two target bearings. Admittedly, it is recognized that the resolution capability of the CBF exhibits certain limitations. Despite MUSIC, SBL and SPICE can achieve high resolution estimation, SPICE experiences a 1dB energy loss in the power spectrum for the target of $20.3^{\circ}$. In contrast, SBL shows a 2dB loss, and the MUSIC method demonstrates a 4.5dB loss. Despite no energy loss in terms of power spectrum, $q$-SPICE experiences slight deviations in both directions due to the fixed grid limitation. Furthermore, both SBL and $q$-SPICE exhibit false peaks in $-47^{\circ}$, $31.5^{\circ}$, $48^{\circ}$, even at extremely low energy levels. The $q$-SPICE-$\text{GNR}^2$ demonstrates superior performance in resolution, target spatial power spectrum energy, and estimation accuracy, while avoiding false peaks.

\begin{table}[!t]
	\caption{The simulation parameters of different SNR and different snapshots with uniform spatial noise setting.\label{tab:Table01}}
	\centering
	\begin{tabular}{|c|c|c|} 
		\hline 
		Parameters &Varied SNR & Varied Snapshot\\
		\hline  
		SNR&-3:3:15&5\\
		
		Snapshots&60&30:30:150\\
		
		True DOAs&[2.36; 27.62]&[2.36; 27.62]\\
		
		The number of elements&12&12\\
		
		Noise Type&spatial uniformly&spatial uniformly\\
		
		The number of Monte Carlo&500&500\\
		\hline 
	\end{tabular}
\end{table}

Fig.\ref{Fig:estDOA_Non_unifNois}a provides the estimated spatial power spectrum generated by all methods for non-uniformly Gaussian spatial noise, employing an arbitrary linear array configuration. Fig.\ref{Fig:estDOA_Non_unifNois}b depicts a detailed result of the DOA region, ranging from $-1^{\circ}$ to $31^{\circ}$. Fig.\ref{Fig:estDOA_Non_unifNois}c also displays the precise results for SBL and $q$-SPICE, aiming to enhance the visibility of spatial spectrum. As present in the Fig.\ref{Fig:estDOA_Non_unifNois}, all methods are capable of estimating two target bearings. Due to aliasing of the spatial power spectrum caused by arbitrary linear arrays, CBF presents a higher energy sidelobe at $-38^{\circ}$, in addition to its lack of resolution. For the target in $2.37^{\circ}$, entire approaches except for the $q$-SPICE-$\text{GNR}^2$ experience energy losses in the power spectrum, with SBL showing the most significant loss at 5dB. Compared to the scenario with uniform noise, both MUSIC and SPICE exhibit a 6dB increase in sidelobe level in non-uniform noise environments. This increase predominantly arises from the failure of these two techniques to suppress and separate non-uniform noise effectively. To conclude, $q$-SPICE-$\text{GNR}^2$ also presents superior sensing performance in challenging environments (non-uniform noise/position errors) comparable to ideal conditions (uniform settings), by enhancing resolution, target spatial power spectrum energy, and sensing accuracy, while effectively mitigating false peaks.

\begin{figure}[!t]
	\centering
	\includegraphics[width=7cm, height=!]{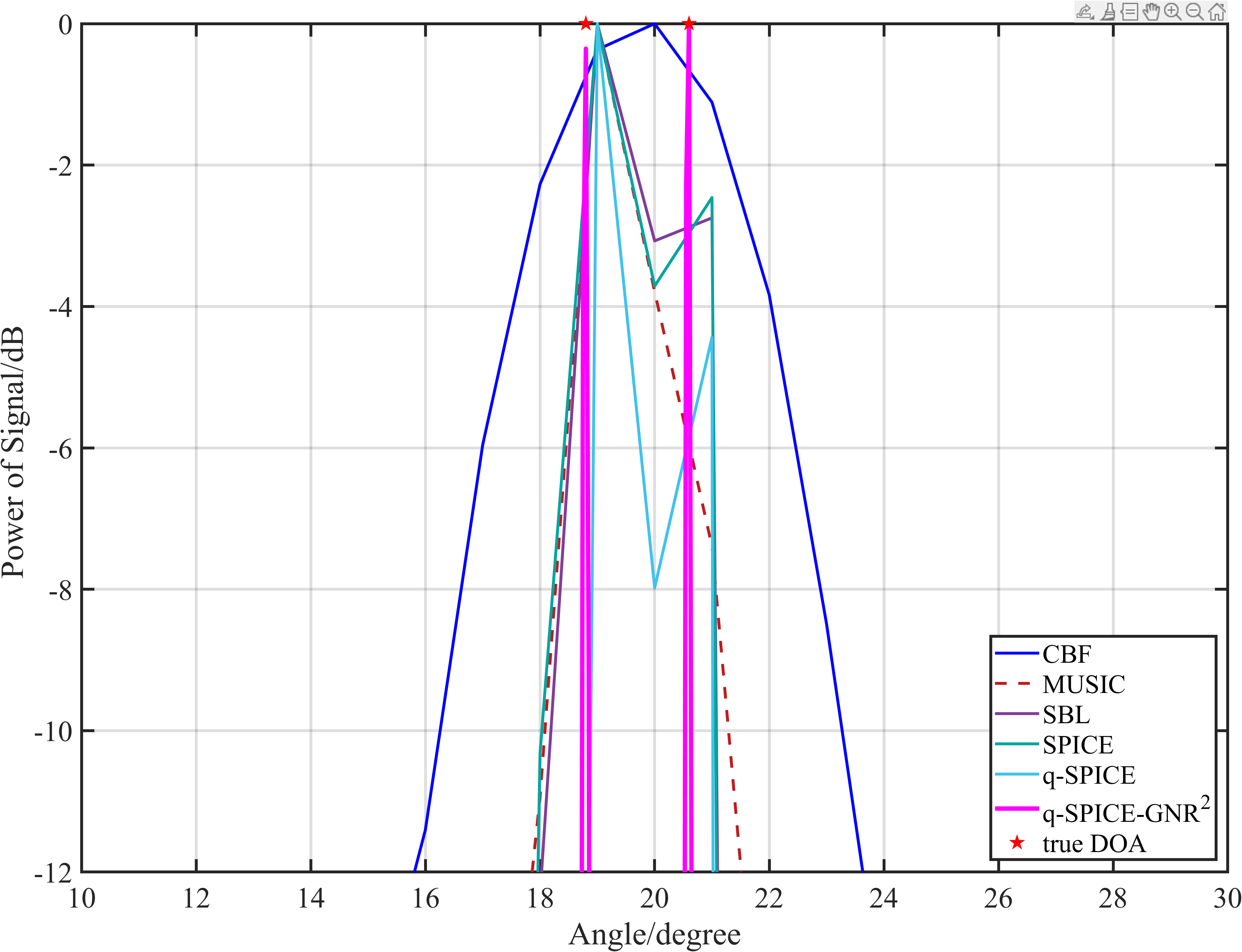}
	\caption{The estimated spatial power spectrum by various methods for two closely spaced broadband propeller noise sources.}
	\label{Fig:closeWide}
	\vspace{-3mm}
\end{figure}

To evaluate the estimation performance of different methods under broadband sources, we simulate broadband propeller radiation noise. As described in the literature\cite{Guo2023Variational}, propeller radiation noise includes both continuous and line spectra, which is suitable and crucial for simulating broadband underwater targets. The frequency range of broadband propeller radiation noise is set from 100 Hz to 1000 Hz. In this case, a non-uniform noise environment is considered, with an SNR of 0 dB. Two closely spaced true azimuths, 18.8° and 20.6°, are defined. As shown in Fig.\ref{Fig:closeWide}, CBF fails to resolve the two targets. Furthermore, other grid-based methods are unable to estimate the azimuths accurately due to limitations imposed by their fixed spatial sampling. In contrast, $q$-SPICE-$\text{GNR}^2$ achieves stable and precise estimation performance.

\begin{figure}[t]
	\centering
	\includegraphics[width=\columnwidth]{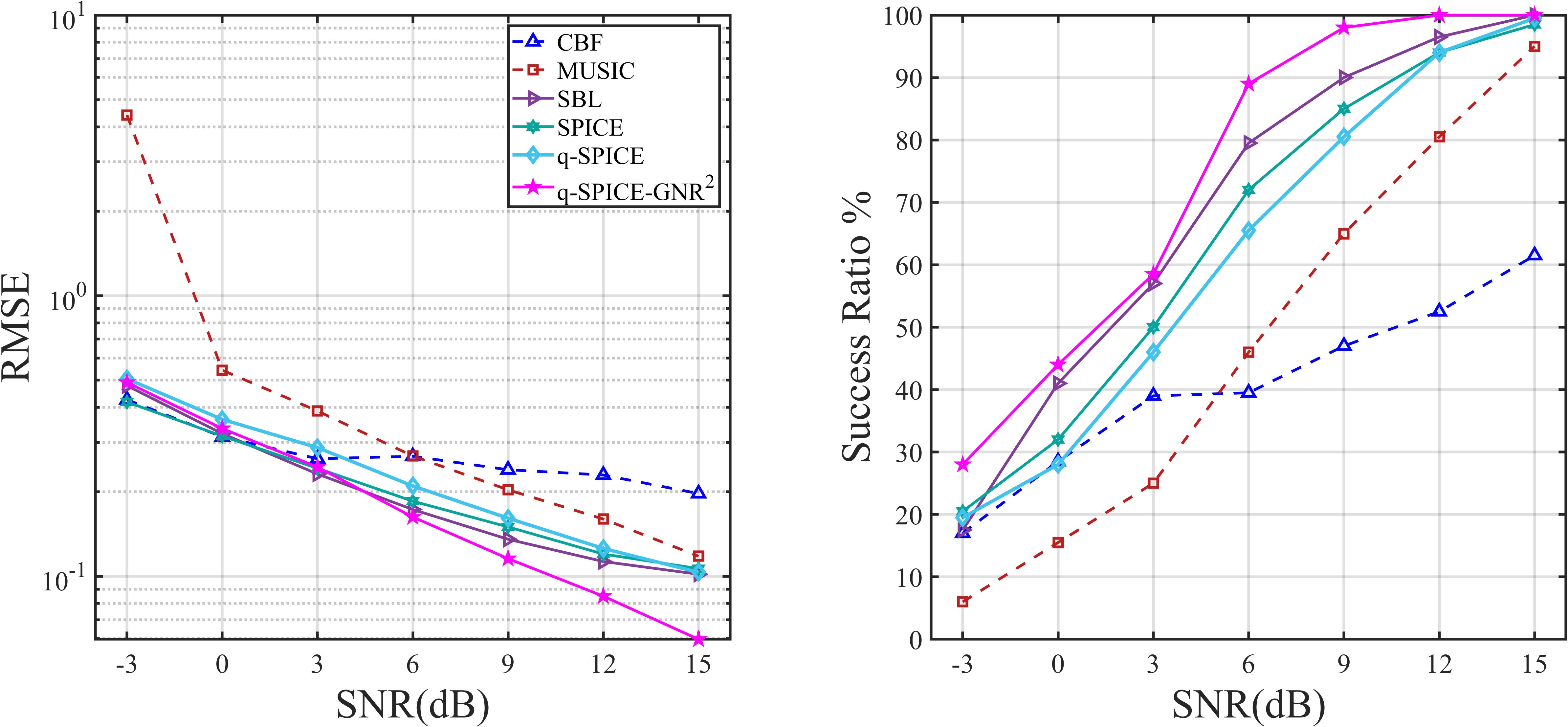}
	\caption{The statistical result for uniform noise with different SNR.}
	\label{Fig:3_StatsResult_unifNois_diffSNR}
	\vspace{-3mm}
\end{figure}

\begin{figure}[t]
	\centering
	\includegraphics[width=\columnwidth]{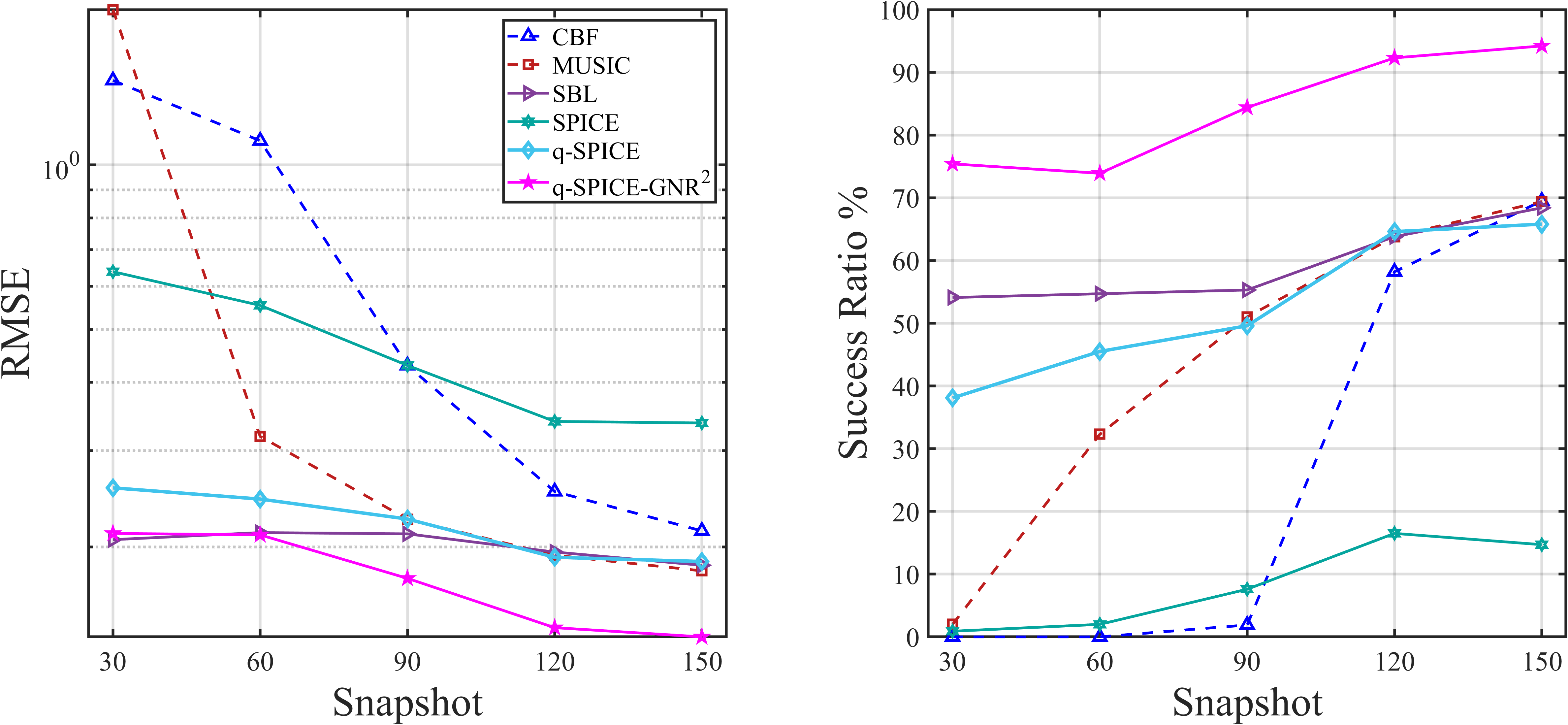}
	\caption{The statistical result for uniform noise with different snapshot.}
	\label{Fig:4_StatsResult_unifNois_diffSnapshot}
	\vspace{-3mm}
\end{figure}
\begin{figure}[t]
	\centering
	\includegraphics[width=\columnwidth]{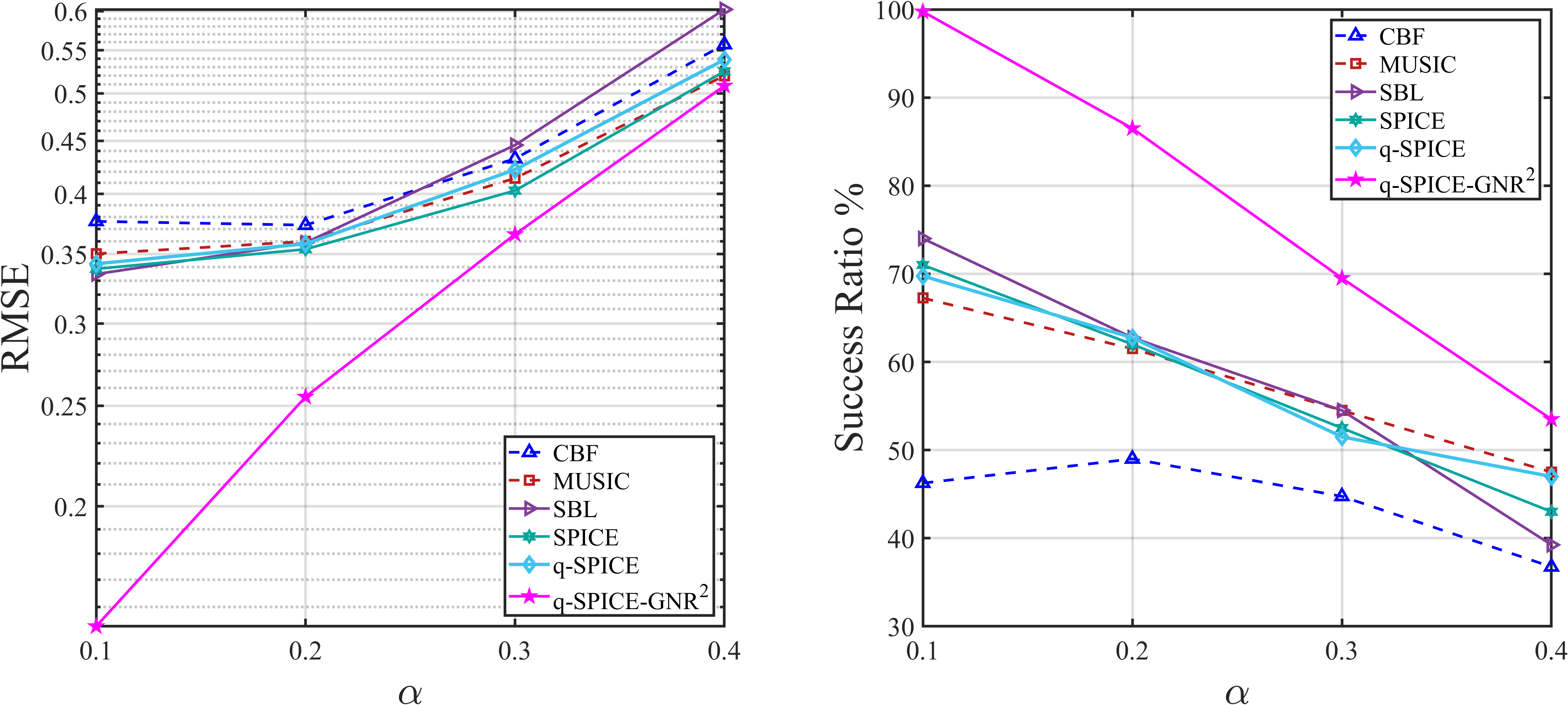}
	\caption{The statistical result for uniform noise with array position error.}
	\label{Fig:5_StatsResult_unifNois_PosError_diffAlpha}
	\vspace{-3mm}
\end{figure}
\begin{figure}[t]
	\centering
	\includegraphics[width=\columnwidth]{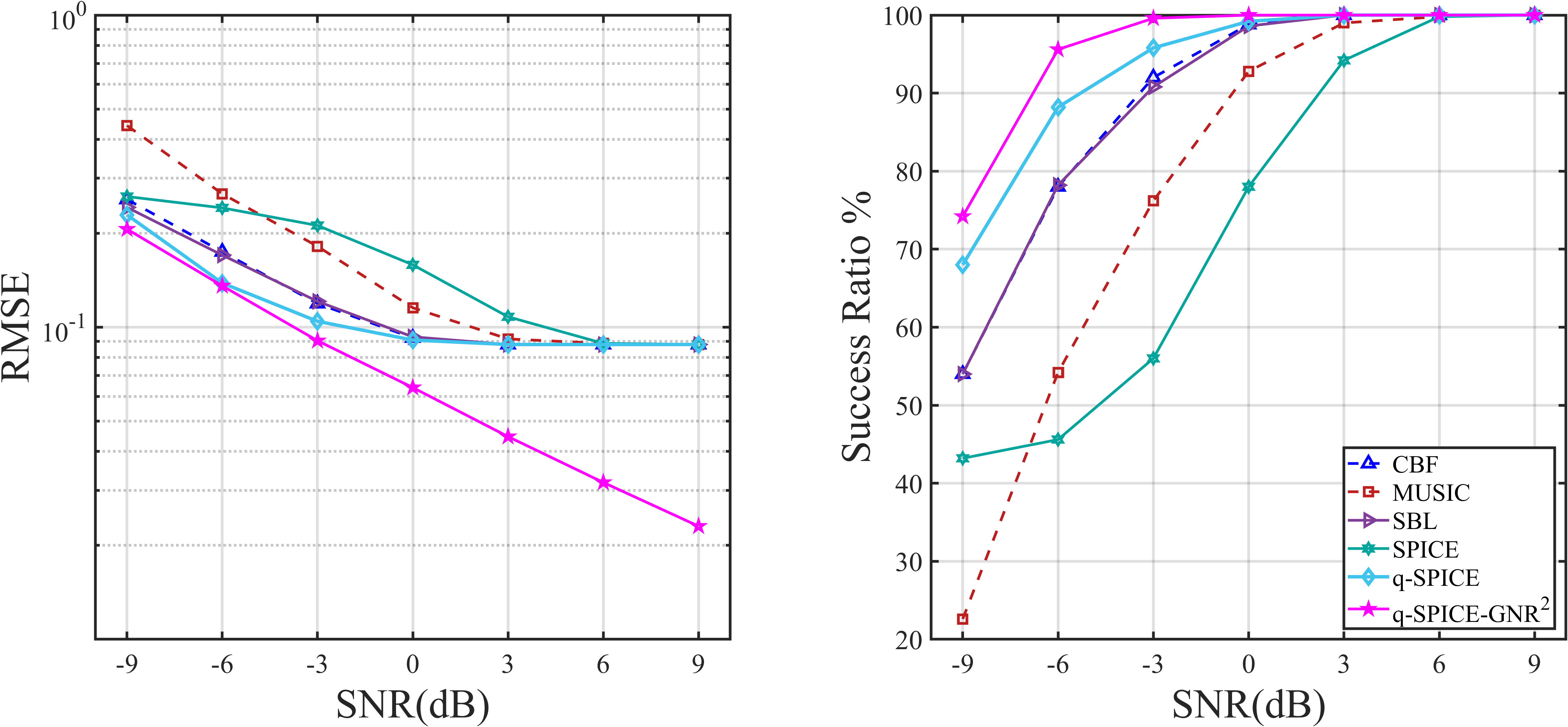}
	\caption{The statistical result for non-uniform noise with different SNR.}
	\label{Fig:6_StatsResult_Non_unifNois_diffSNR}
	\vspace{-3mm}
\end{figure}
\subsection{The performance comparison under uniform noise}
We herein clarify the concept of ‘success’. Each estimate of DOAs will be compared to the true value. If the absolute errors of all estimates are smaller than the threshold, the estimation is deemed as ‘success’. In this article, the success ratio threshold is defined as 0.3 degrees. Thus, the success ratio can be calculated by
\begin{equation}
	\text{success ratio} \%  =\frac{N_{suces}}{M_{c}} \times 100
\end{equation}
where $N_{suces}$ represents the total number of success, $M_{c}$ denotes the total number of Monte Carlo running. The Root Mean Squared Error (RMSE) of the DOA estimates is defined as 
\begin{equation}
	\text{RMSE}=\sqrt{\frac1{M_c K}\sum_{\nu=1}^{M_c} \sum_{k=1}^K{(\hat{\theta}_{k,\nu}-\theta_k)^2}}
\end{equation}
where $\hat{\theta}_{k,\nu}$ denotes the estimated DOA in the $\nu$-th simaltion for $k$-th source. $\theta_k$ presents the corresponding ground truth. We consider $M=12$ elements with half-wavelength spacing. Assuming the presence of two far-field narrowband signal sources, each arriving at the ULA from angles of $\theta_1= 2.36$ and $\theta_2= 27.62$. We evaluate the performance across different SNR and varying numbers of snapshots. The parameter settings are listed in TABLE~\ref{tab:Table01}, with SNR ranging from -3dB to 15dB in 3dB intervals, and snapshot ranging from 30 to 150 in 30 intervals. The RMSE is averaged from 500 trials. Fig.\ref{Fig:3_StatsResult_unifNois_diffSNR} illustrates that SPICE attains lower RMSE at low SNR levels, while $q$-SPICE-$\text{GNR}^2$ exhibits improved performance as the SNR increases. In Fig.\ref{Fig:4_StatsResult_unifNois_diffSnapshot}, it is observed that $q$-SPICE-$\text{GNR}^2$ can be seen to yield results quite similar to SBL when the snapshots are configured at 30 and 60. As the number of snapshots increases, GNR shows a gradual improvement in performance. By analyzing Fig.\ref{Fig:3_StatsResult_unifNois_diffSNR} and Fig.\ref{Fig:4_StatsResult_unifNois_diffSnapshot} together, it is apparent that $q$-SPICE-$\text{GNR}^2$ has been leading in success ratio, demonstrating excellent performance in sensing resolution for target bearing.
\begin{table}[!t]
	\caption{The simulation parameters of different level of position errors with uniform spatial noise setting.\label{tab:Table02}}
	\centering
	\begin{tabular}{|c|c|} 
		\hline 
		Parameters& Values\\
		\hline  
		SNR&10\\
		
		Snapshots&60\\
		
		Array Position Error Level& 0.1:0.1:0.4\\
		
		True DOAs&[-2.36; 19.78]\\
		
		The number of elements&12\\
		
		Noise Type&spatial uniformly\\
		
		The number of Monte Carlo&500\\
		\hline
	\end{tabular}
\end{table}

Additionally, we assess performance by introducing errors in array positions. TABLE~\ref{tab:Table02} provides a comprehensive list of the general parameters. Noting that the array position error varies within the range of $0.1 * d$ to $0.4 * d$, with intervals of $0.1 * d$. The test statistics, along with the RMSE and success ratio for all methods, are illustrated in Fig.\ref{Fig:5_StatsResult_unifNois_PosError_diffAlpha}. In cases where the array position error is around $0.1*d$, $q$-SPICE-$\text{GNR}^2$ outperforms alternative methods. However, as the error increases, the efficiency of all techniques diminishes notably, albeit with $q$-SPICE-$\text{GNR}^2$ retaining a slight lead. This suggests that $q$-SPICE-$\text{GNR}^2$ exhibits a level of resilience against substantial array position errors.
\begin{figure}[t]
	\centering
	\includegraphics[width=\columnwidth]{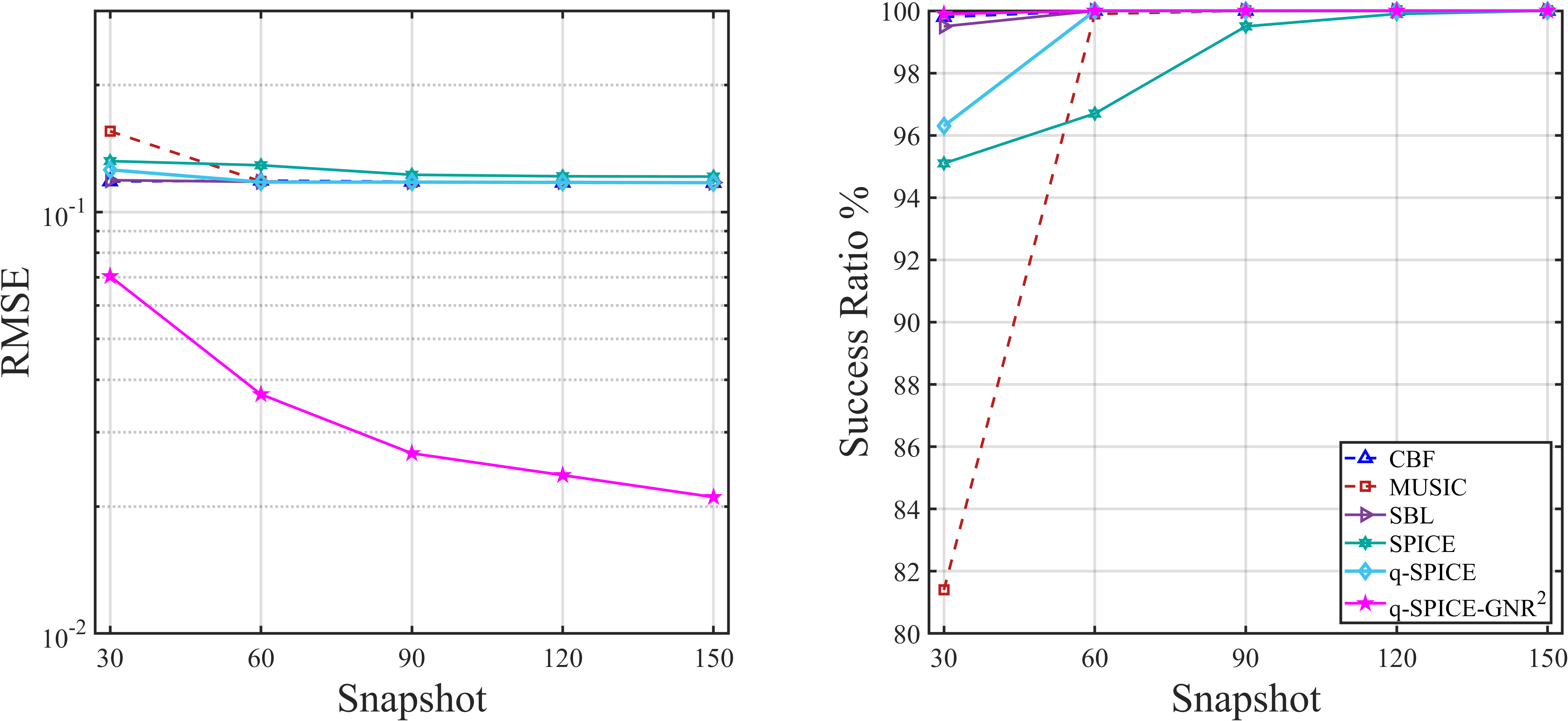}
	\caption{The statistical result for non-uniform noise with different snapshot.}
	\label{Fig:7_StatsResult_Non_unifNois_diffSnapshot}
	\vspace{-3mm}
\end{figure}
\begin{figure}[t]
	\centering
	\includegraphics[width=\columnwidth]{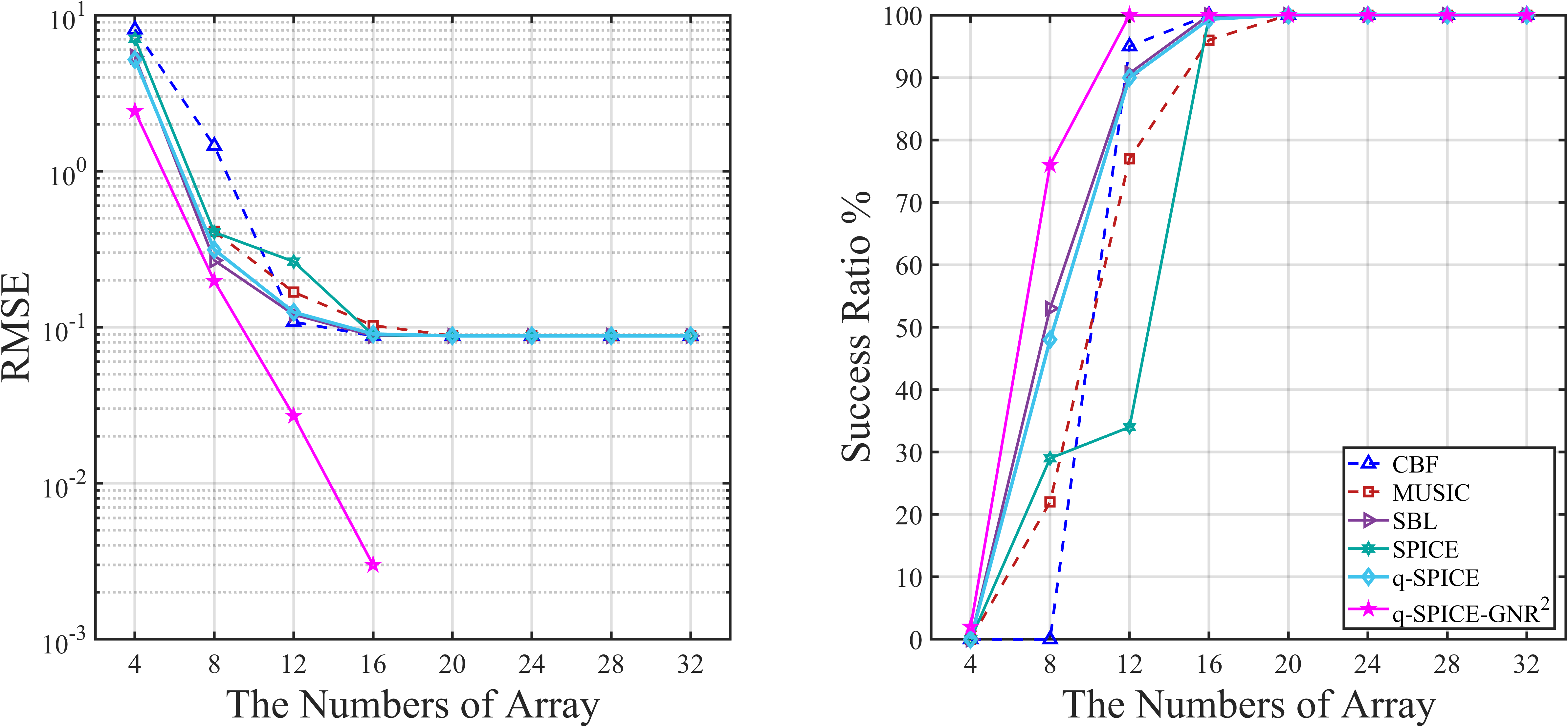}
	\caption{The statistical result for non-uniform noise with different number of array element $M$.}
	\label{Fig:8_StatsResult_Non_unifNois_diffM}
	\vspace{-3mm}
\end{figure}
\begin{figure}[t]
	\centering
	\includegraphics[width=\columnwidth]{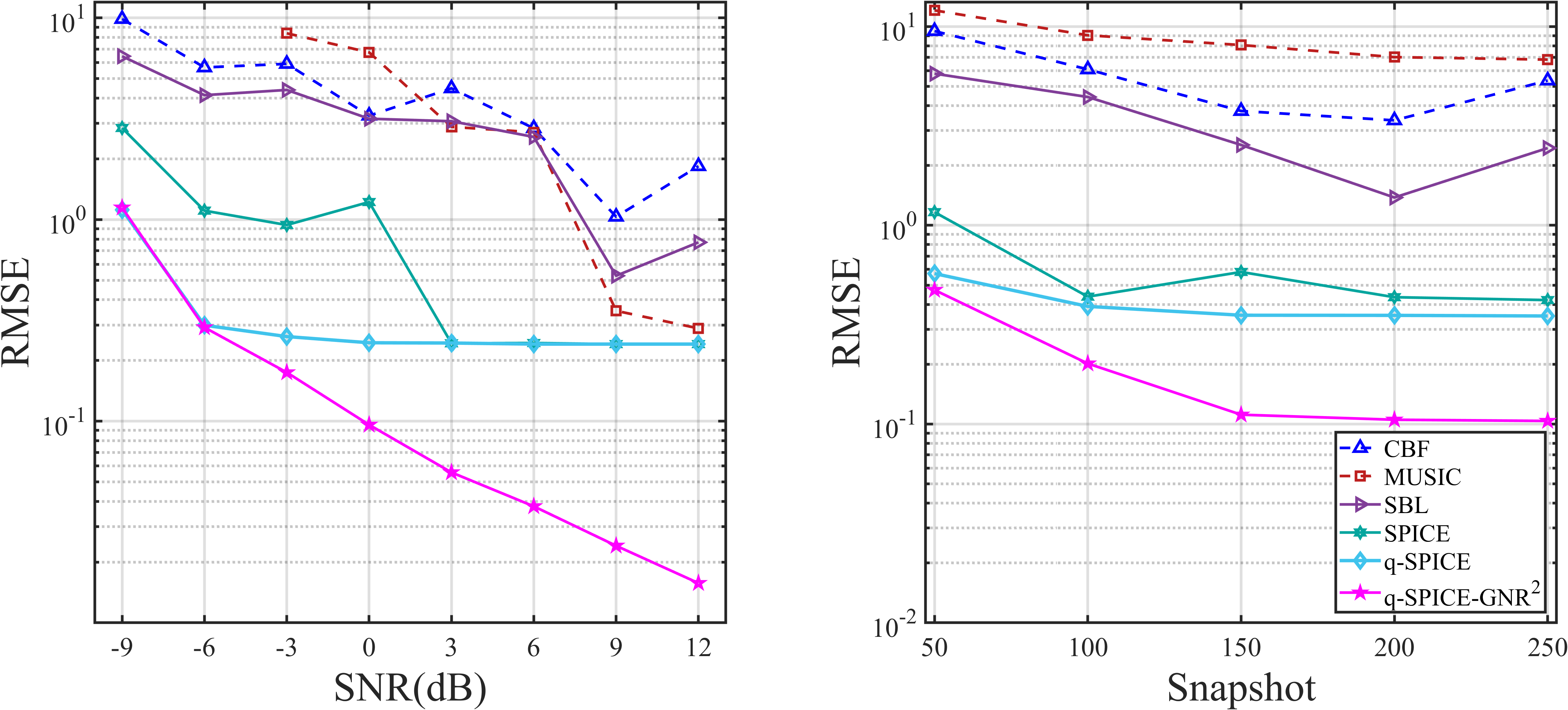}
	\caption{The statistical result for impulsive noise with different SNR and varying snapshot.}
	\label{Fig:21_StatsResult_ImpulsiveNois}
	\vspace{-3mm}
\end{figure}
\subsection{The performance comparison under non-uniform noise} \label{Sec4.3}
Non-uniform spatial noise is incorporated at this stage to examine the outcomes, with the relevant parameters outlined in TABLE \ref{tab:Table03}. The main diagonal element of the noise covariance matrix is defined as [12, 2.3, 20.5, 5.5, 11.1, 6.5, 2, 13.5, 0.8, 1.7, 13.6, 5.2]. The SNR is initially varied from -9dB to 9dB, with the snapshot held constant at 60. Following this, the number of snapshots is altered from 30 to 150, maintaining a fixed non-uniform SNR of -3dB. Figs.\ref{Fig:6_StatsResult_Non_unifNois_diffSNR} and \ref{Fig:7_StatsResult_Non_unifNois_diffSnapshot} illustrate the RMSE and the success ratio curve as they vary with SNR and the number of snapshots. As depicted, $q$-SPICE-$\text{GNR}^2$ exhibits robust sensing capabilities in comparison to alternative methods, regardless of changing SNR or snapshots. The efficacy of the original SPICE method is significantly declining. Because it does not consider to separate the signal subspace and noise subspace by different norm constraint.
\begin{table}[!t]
	\caption{The simulation parameters of different SNR and different snapshots with non-uniform spatial noise setting.\label{tab:Table03}}
	\centering
	\begin{tabular}{|c|c|c|} 
		\hline 
		Parameters& Varied SNR &  Varied Snapshot\\
		\hline  
		SNR&-9:3:9&-3\\
		
		Snapshots&60&30:30:150\\
		
		True DOAs&[2.36; 27.62]&[2.36; 27.62]\\
		
		The number of elements&12&12\\
		
		Noise Type&non-uniform&non-uniform\\
		
		The number of Monte Carlo&500&500\\
		\hline
	\end{tabular}
\end{table}

We evaluate the performance across different numbers of array elements, while keeping the non-uniform SNR and snapshot fixed at 0dB and 60. The main diagonal element of the noise covariance matrix is generated by the form of $ 0.5 + 25 * rand(1,M)$. The constant 0.5 is intentionally introduced as a baseline noise floor. This assumption prevents the noise variance from approaching zero, thereby avoiding numerical instability. And the upper limit of 25.5 represents a reasonable maximum noise level for the simulation setup. The array elements range from 4 to 32, with intervals of 4 elements and a half-wavelength configuration. The comparison results depicted in Fig.\ref{Fig:8_StatsResult_Non_unifNois_diffM} reveal that $q$-SPICE and $q$-SPICE-$\text{GNR}^2$ offer slight leads using fewer elements. But as the array aperture expands, $q$-SPICE-$\text{GNR}^2$ demonstrates a substantial superiority. Once the array comprises more than 16 elements, $q$-SPICE-$\text{GNR}^2$ eliminates estimation errors.

\begin{figure}[t]
	\centering
	\includegraphics[width=\columnwidth]{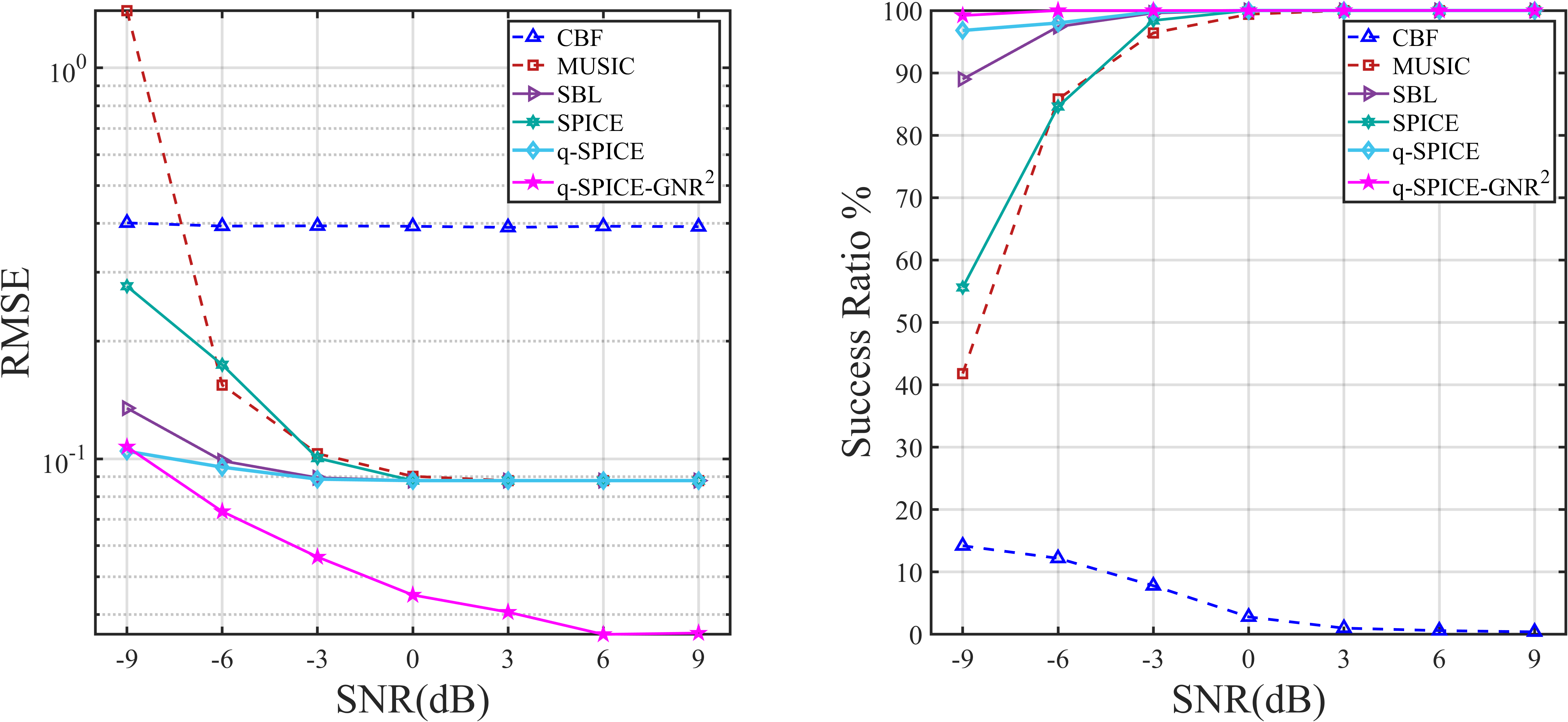}
	\caption{The statistical result for non-uniform noise and element spacing errors with different SNR.}
	\label{Fig:9_StatsResult_Non_unifNoisAndArray_diffSNR}
	\vspace{-3mm}
\end{figure}
\begin{figure}[t]
	\centering
	\includegraphics[width=7cm, height=!]{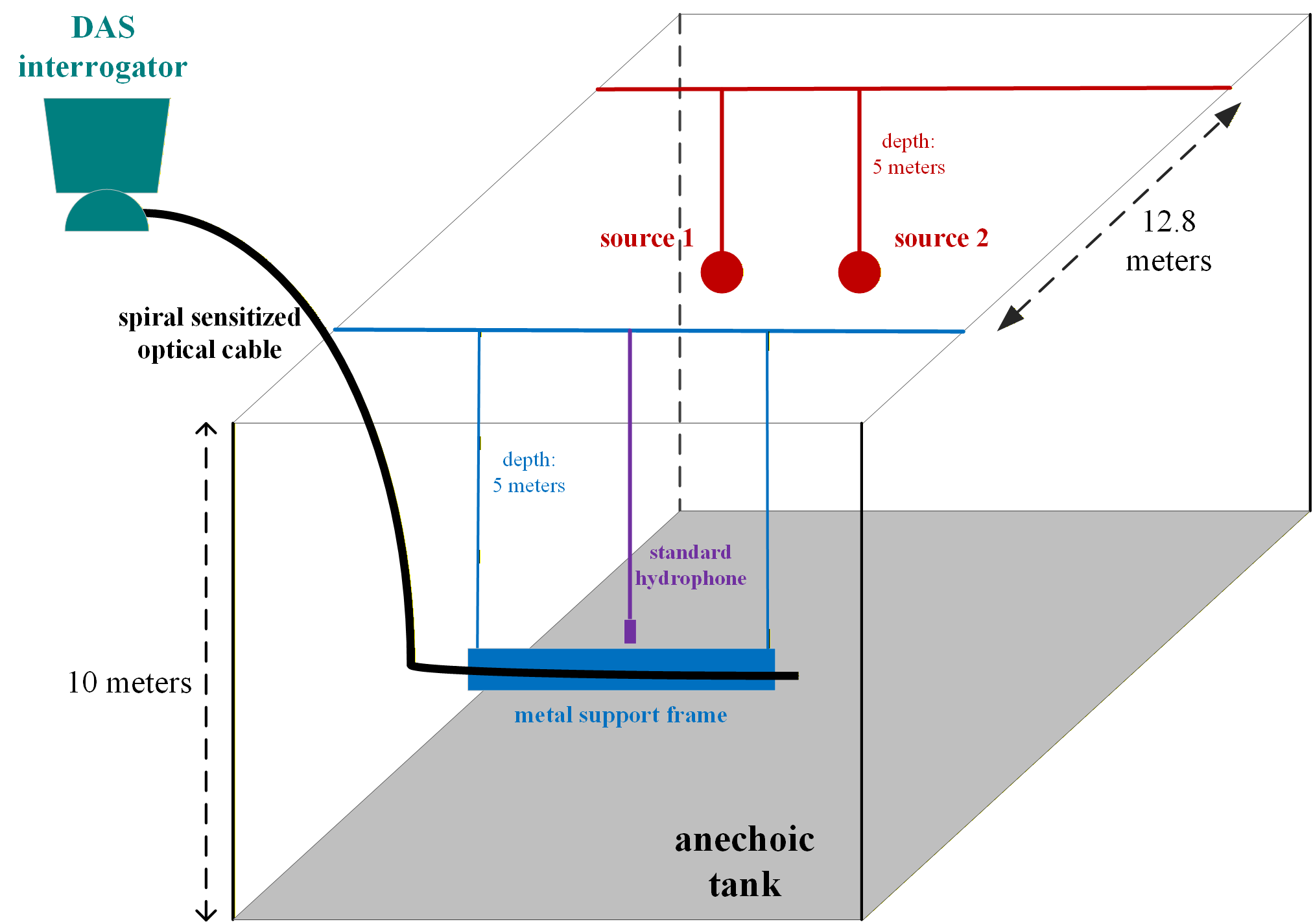}
	\caption{The schematic diagram of the tank experimental setup.}
	\label{Fig:tank_deploy}
	\vspace{-3mm}
\end{figure}
\begin{figure}[h]
	\centering
	\includegraphics[width=7cm, height=!]{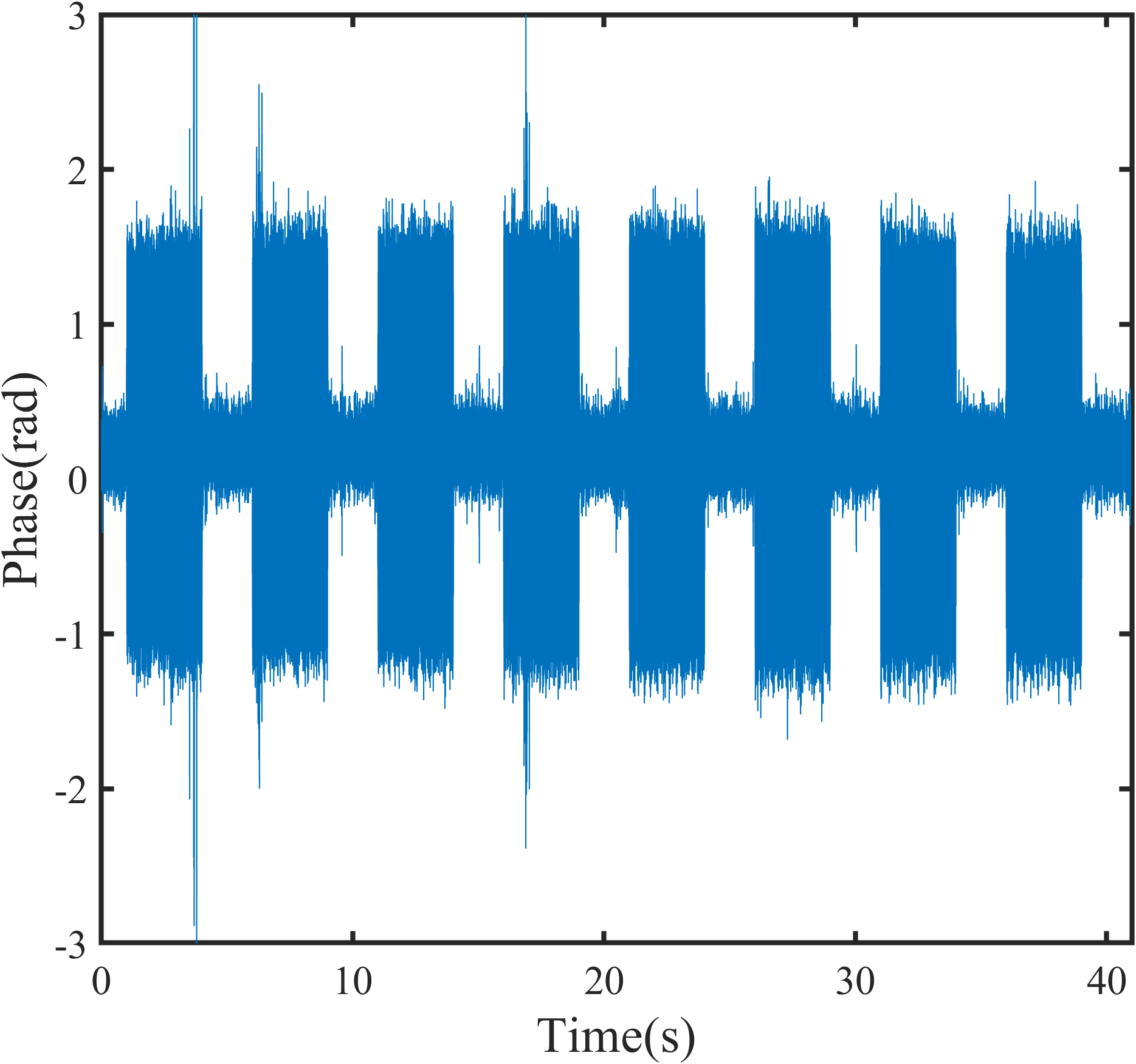}
	\caption{The waveform of DAS signal in the tank.}
	\label{Fig:TankDASwaveform}
	\vspace{-3mm}
\end{figure}

\subsection{The performance comparison under impulsive noise}
In practical underwater environments, there exists a significant amount of impulsive noise; therefore, performance tests are also conducted under impulsive noise condition. The relevant simulation parameters are presented in TABLE~\ref{tab:Table07}. For the generated impulsive noise, the S$\alpha$S model was used, with four typical parameters set as $\alpha=1.2, \beta=0, \delta=0, \gamma=1$. Fig.\ref{Fig:21_StatsResult_ImpulsiveNois} illustrates the RMSE statistics under different SNR levels and varying snapshot counts. It is worth mentioning that the $q$-SPICE-$\text{GNR}^2$ showed reliable estimation performance despite the existence of impulsive noise. Variations in impulsive noise energy distribution across different channels are thought to be significant, similar to the non-uniform noise characteristics discussed previously. Consequently, the estimation results obtained through the $q$-SPICE-$\text{GNR}^2$ method are comparable to those detailed in the Section \ref*{Sec4.3}. 
\begin{table}[!t]
	\caption{The simulation parameters of different SNR and different snapshots with impulsive noise condition.\label{tab:Table07}}
	\centering
	\begin{tabular}{|c|c|c|} 
		\hline 
		Parameters& Varied SNR &  Varied Snapshot\\
		\hline  
		SNR&-9:3:12&0\\
		
		Snapshots&400&50:50:250\\
		
		True DOAs&[2.36; 27.62]&[2.36; 27.62]\\
		
		The number of elements&12&12\\
		
		Noise Type&impulsive&impulsive\\
		
		Noise model&S$\alpha$S&S$\alpha$S\\
		
		The number of Monte Carlo&500&500\\
		\hline
	\end{tabular}
\end{table}

\subsection{The performance comparison under the element spacing errors}
\begin{table}[!t]
	\caption{The simulation parameters of different SNR with non-uniform spatial noise and array position errors.\label{tab:Table04}}
	\centering
	\begin{tabular}{|c|c|} 
		\hline 
		Parameters& Values \\
		\hline  
		SNR&-9:3:9 \\
		
		Snapshots&60\\
		
		True DOAs&[2.36; 27.62]\\
		
		The number of elements&12\\
		
		Noise Type&spatial non-uniformly\\
		
		Array Position Error Level & 0.1\\
		
		The number of Monte Carlo&500\\
		\hline
	\end{tabular}
\end{table}
In the last simulation, we then compare the estimating performance both non-uniform noise and the element spacing errors. Underwater topography often causes positional errors in arrays deployed in the water, resulting in deviations from the ideal uniform linear array configuration. The spatial noise is set in a non-uniform manner, closely mimicking the actual underwater remote sensing environment. The configuration of the simulation is set as same as section \ref{section4.1}. The remaining parameters are detailed in TABLE \ref{tab:Table04}, while the outcomes for RMSE and success ratio can be observed in Fig.\ref{Fig:9_StatsResult_Non_unifNoisAndArray_diffSNR}. When considering the element spacing errors, it is evident that strong non-uniform noise with an SNR of -9dB can result in considerable estimation errors when utilizing conventional methods. In contrast, both $q$-SPICE and $q$-SPICE-$\text{GNR}^2$ demonstrate robustness in the face of such challenges. Despite the severe configurations, the $q$-SPICE-$\text{GNR}^2$ algorithm offers precise DOA estimates through noise separating constraint and grid refinement framework. In brief, after comparing Fig.\ref{Fig:3_StatsResult_unifNois_diffSNR}, Fig.\ref{Fig:6_StatsResult_Non_unifNois_diffSNR} and Fig.\ref{Fig:9_StatsResult_Non_unifNoisAndArray_diffSNR} together, it is evident that $q$-SPICE-$\text{GNR}^2$ exhibits superior performance in the presence of non-uniform background noise, particularly under the element spacing errors.	

\section{Experimental Data Analysis} \label{Sec5}
\subsection{Tank Experiment}
Notably, the DAS sensor does not strictly qualify as a conventional underwater acoustic sensor, such as a standard hydrophone. To quantitatively evaluate the ability of DAS system to capture underwater acoustic signal waveforms, we conducted a measurement experiment in a standardized anechoic tank. Fig. \ref{Fig:tank_deploy} displays the schematic diagram of the experimental situation. Two transducers were used as sound sources, with the spiral-sensitized optical cable mounted along a horizontal steel bracket to maintain its linear configuration. The distance between transmission and reception was 12.8 meters, and both the sound sources and DAS cable were placed at a depth of 5 meters. A standard hydrophone was positioned at the center of the DAS array to serve as a reference for comparison.
\begin{figure}[t]
	\centering
	\includegraphics[width=6.5cm, height=!]{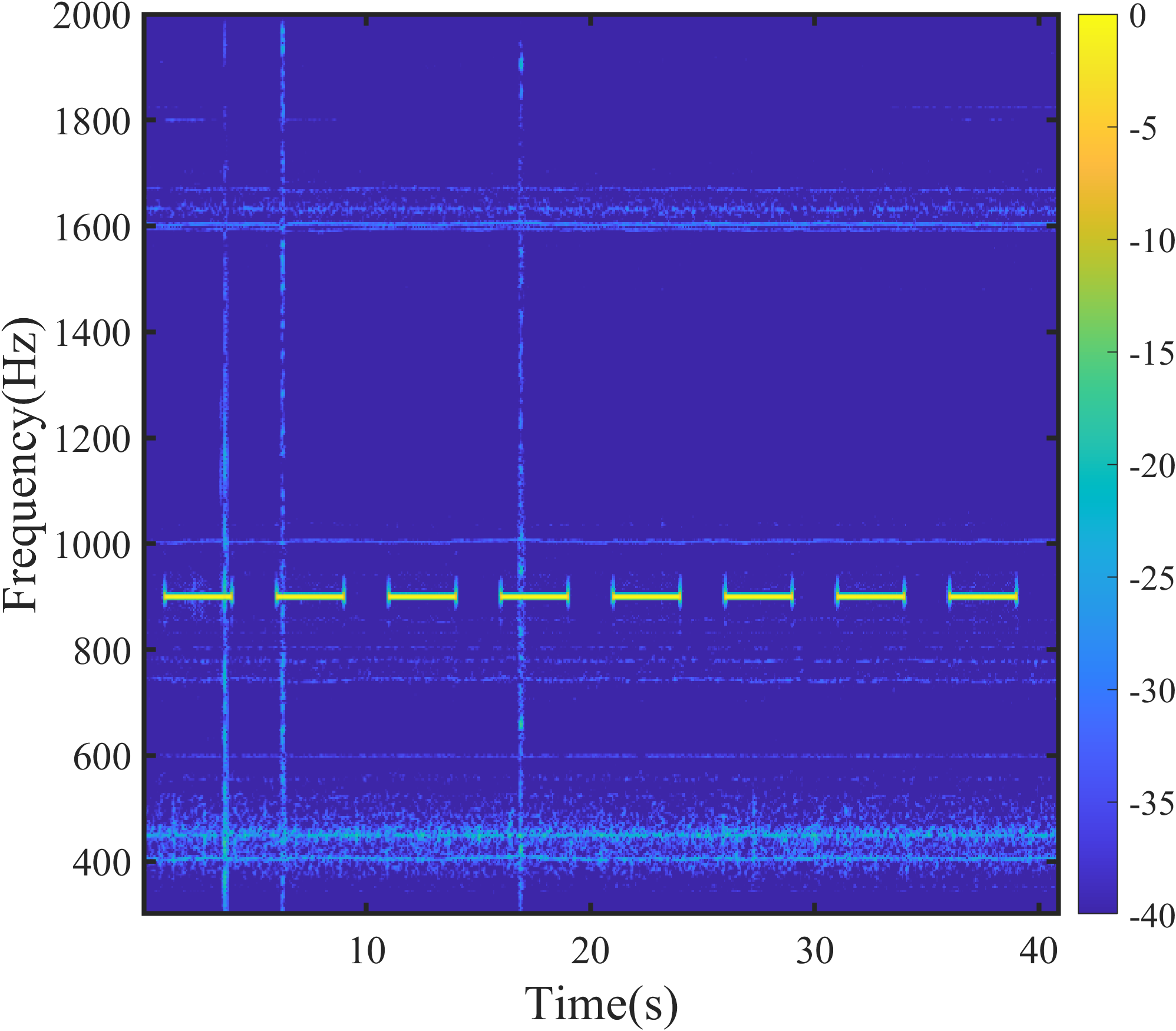}
	\caption{The time-frequency analysis of DAS signal in the tank.}
	\label{Fig:TankDASTF}
	\vspace{-3mm}
\end{figure}

\begin{figure}[h]
	\centering
	\includegraphics[width=6cm, height=!]{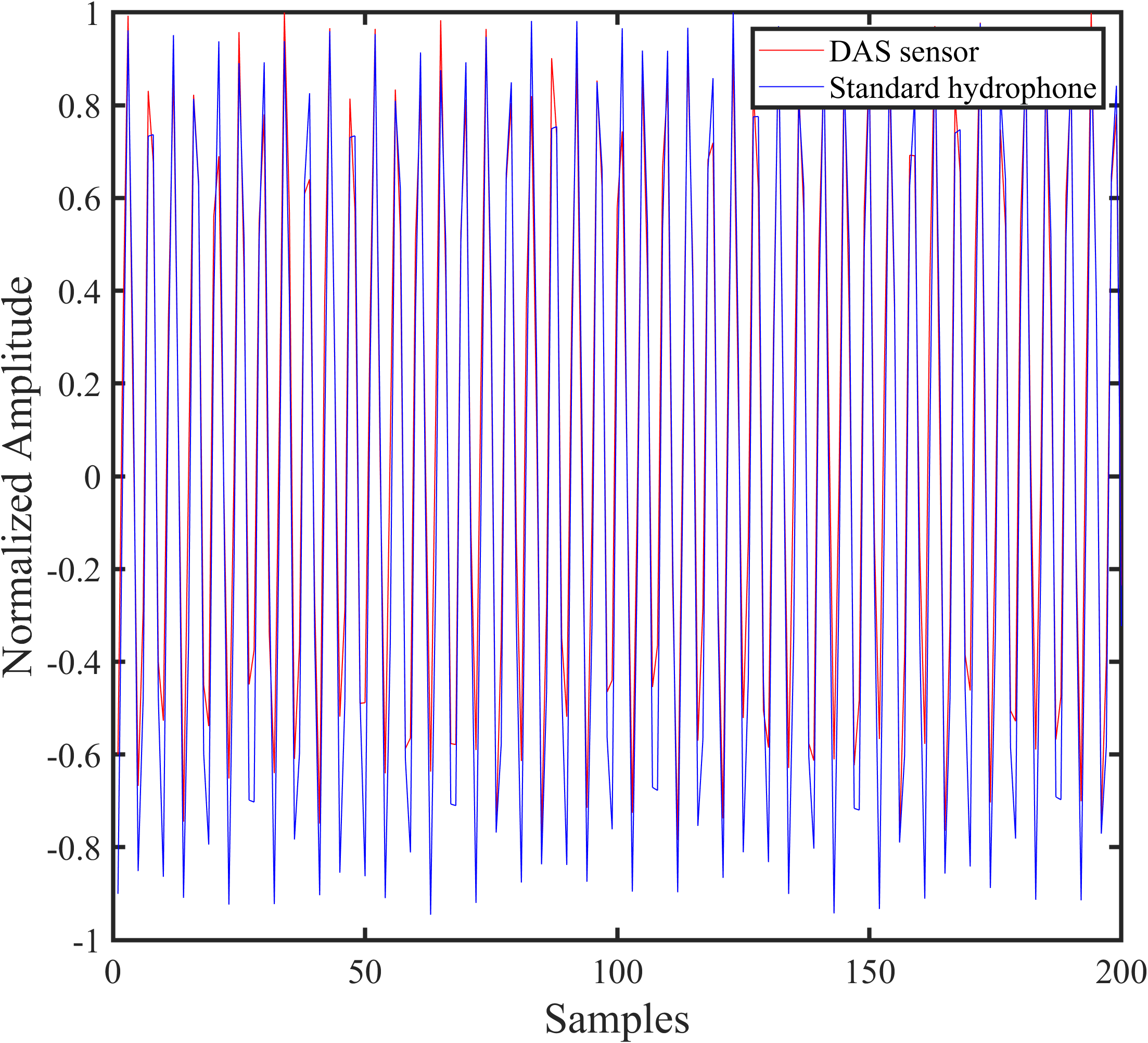}
	\caption{The waveform comparison between signals collected by DAS sensor and standard hydrophone.}
	\label{Fig:CompareDAS}
	\vspace{-3mm}
\end{figure}
\begin{figure}[t]
	\centering
	\includegraphics[width=7cm, height=!]{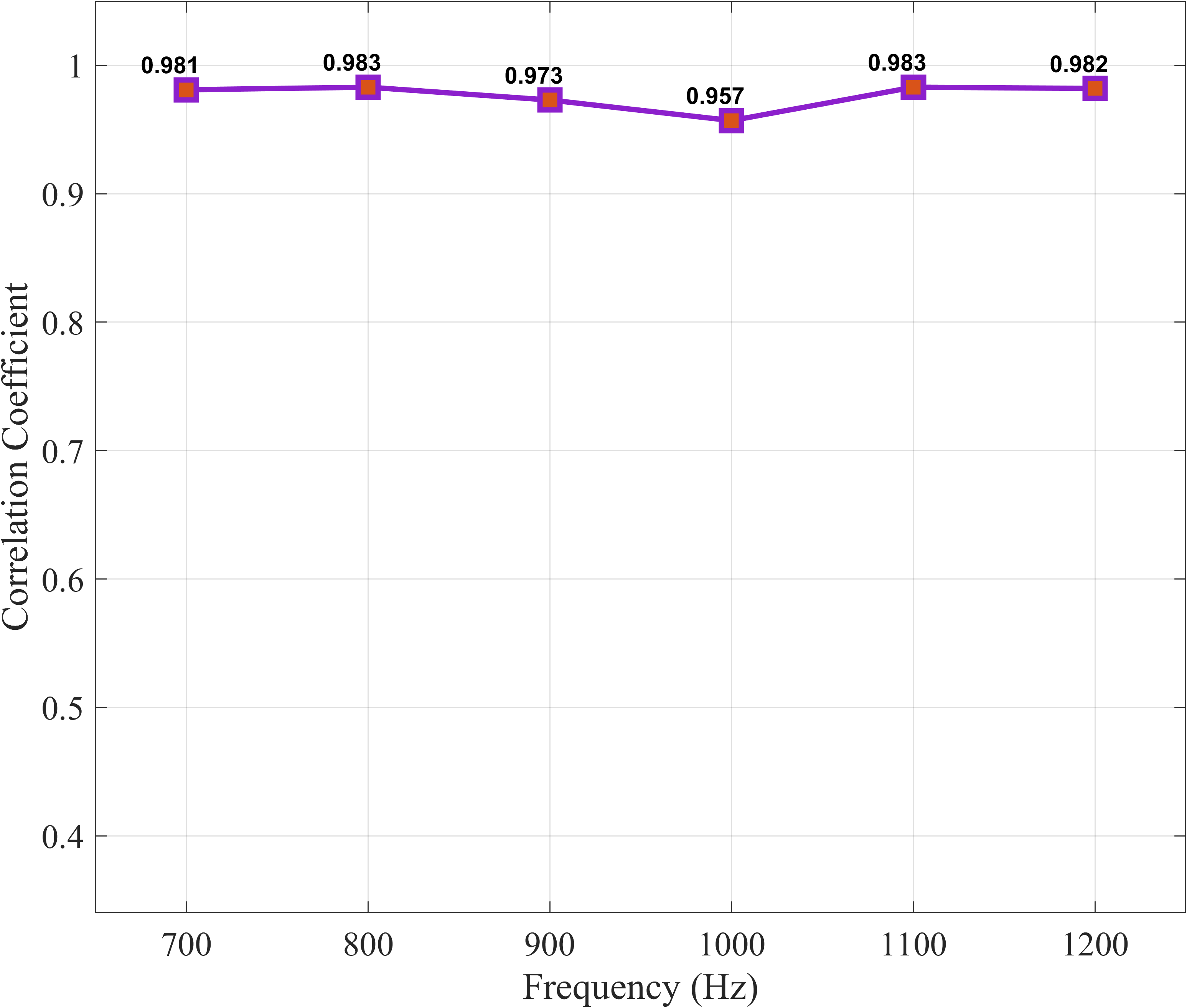}
	\caption{Normalized cross-correlation coefficient comparing the waveform acquired by the DAS system with that from a standard hydrophone.}
	\label{Fig:CC}
	\vspace{-3mm}
\end{figure}
\begin{figure}[t]
	\centering
	\includegraphics[width=7cm, height=!]{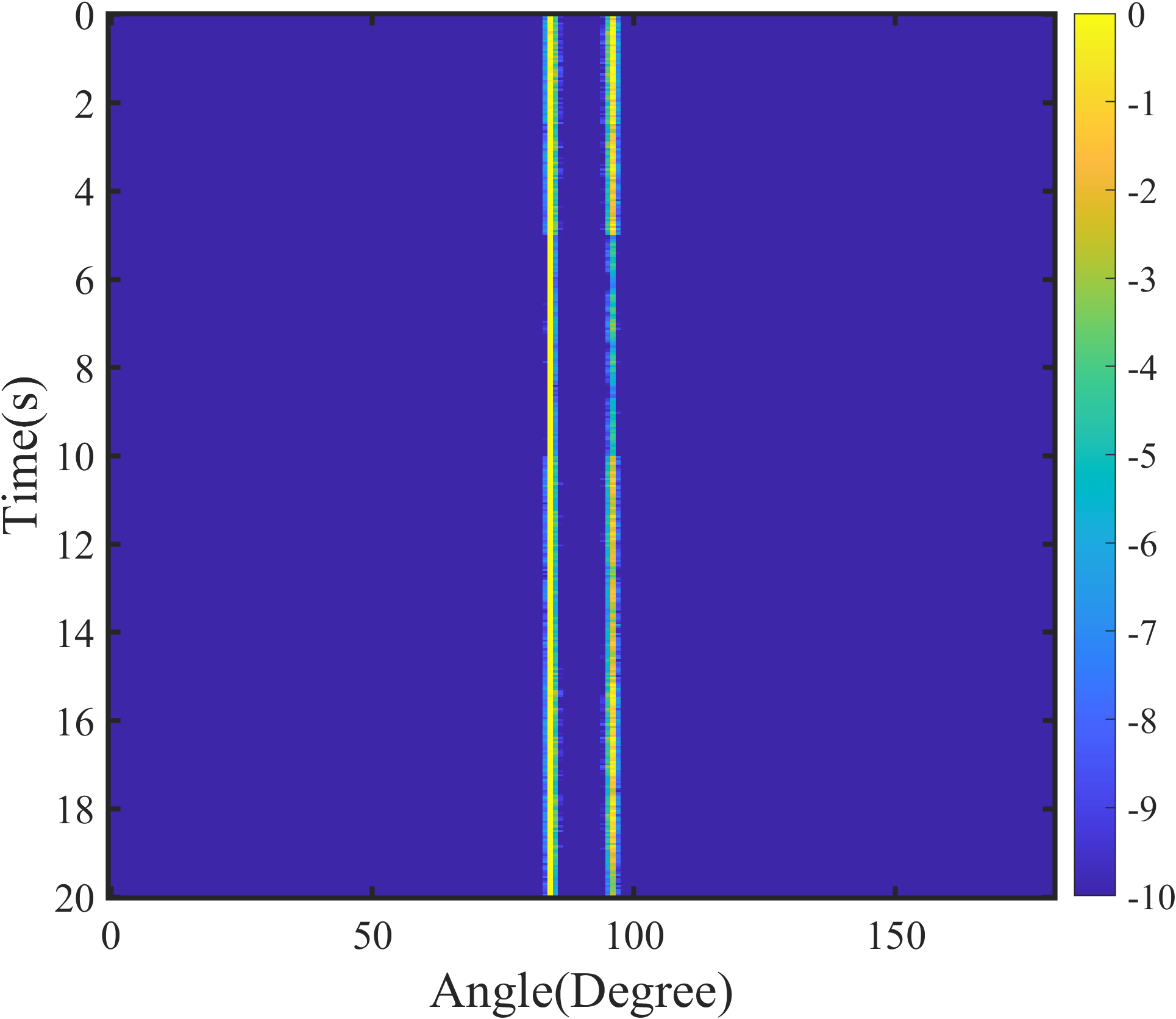}
	\caption{The bearing time record of two sound sources estimated by the $q$-SPICE method.}
	\label{Fig:DAStankBTR}
	\vspace{-3mm}
\end{figure}
\begin{figure}[h]
	\centering
	\includegraphics[width=7cm, height=!]{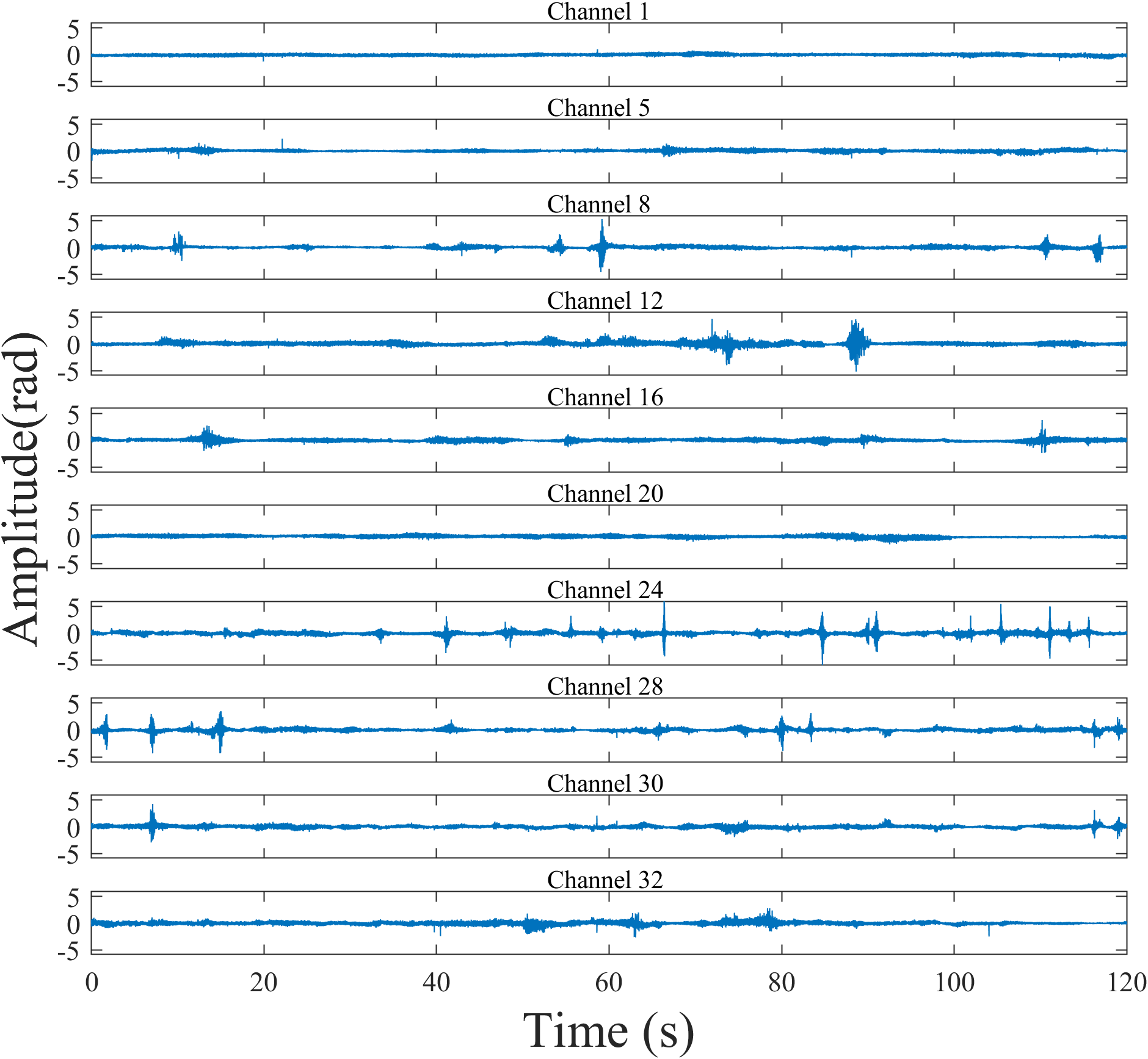}
	\caption{The waveforms of ambient noise recorded across different DAS channels in the lake.}
	\label{Fig:32waveforms}
	\vspace{-3mm}
\end{figure}
\begin{figure}[h]
	\centering
	\includegraphics[width=7cm, height=!]{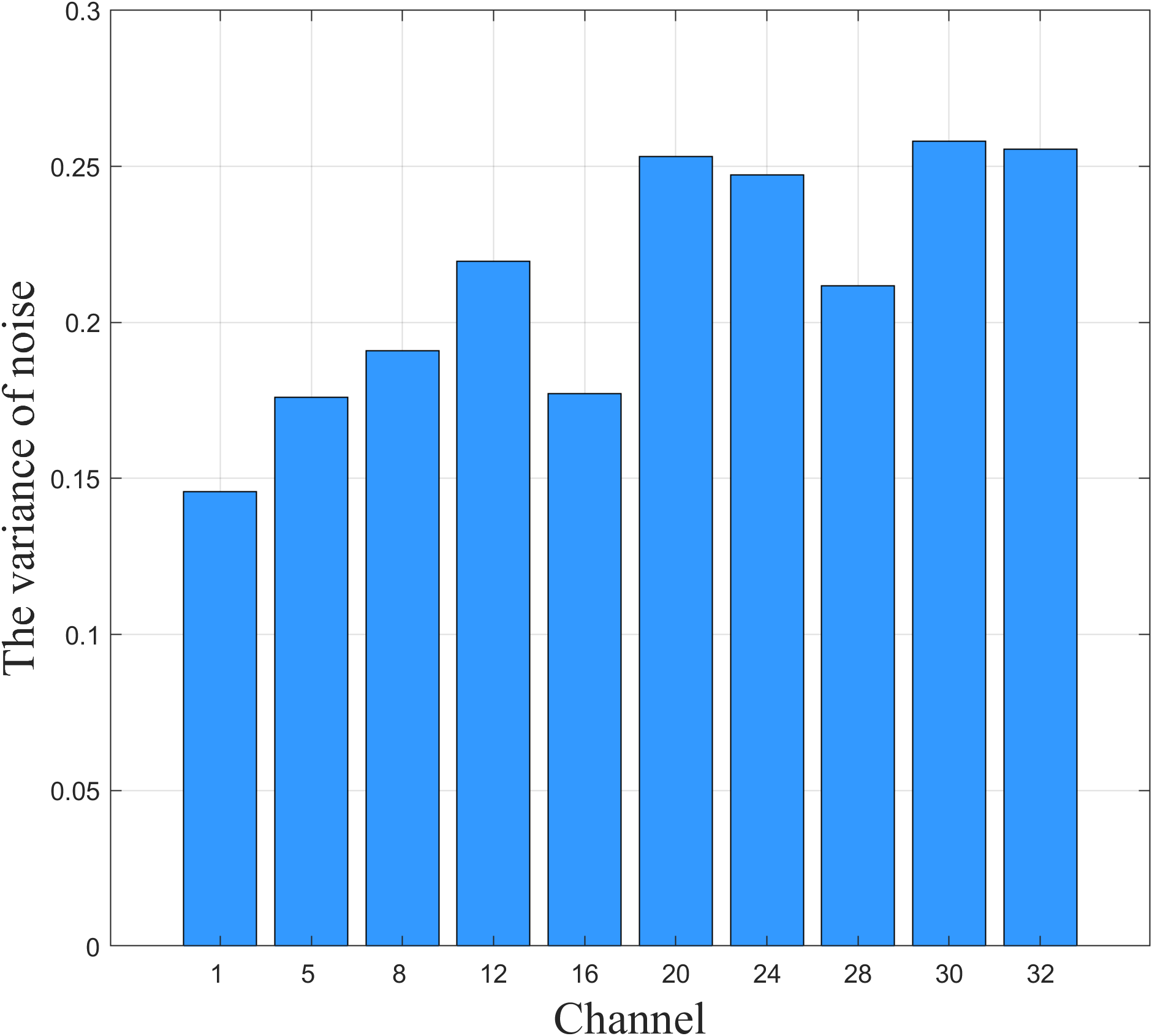}
	\caption{The noise energy distribution across DAS channels.}
	\label{Fig:noise}
	\vspace{-3mm}
\end{figure}
\begin{figure}[h]
	\centering
	\includegraphics[width=7cm, height=!]{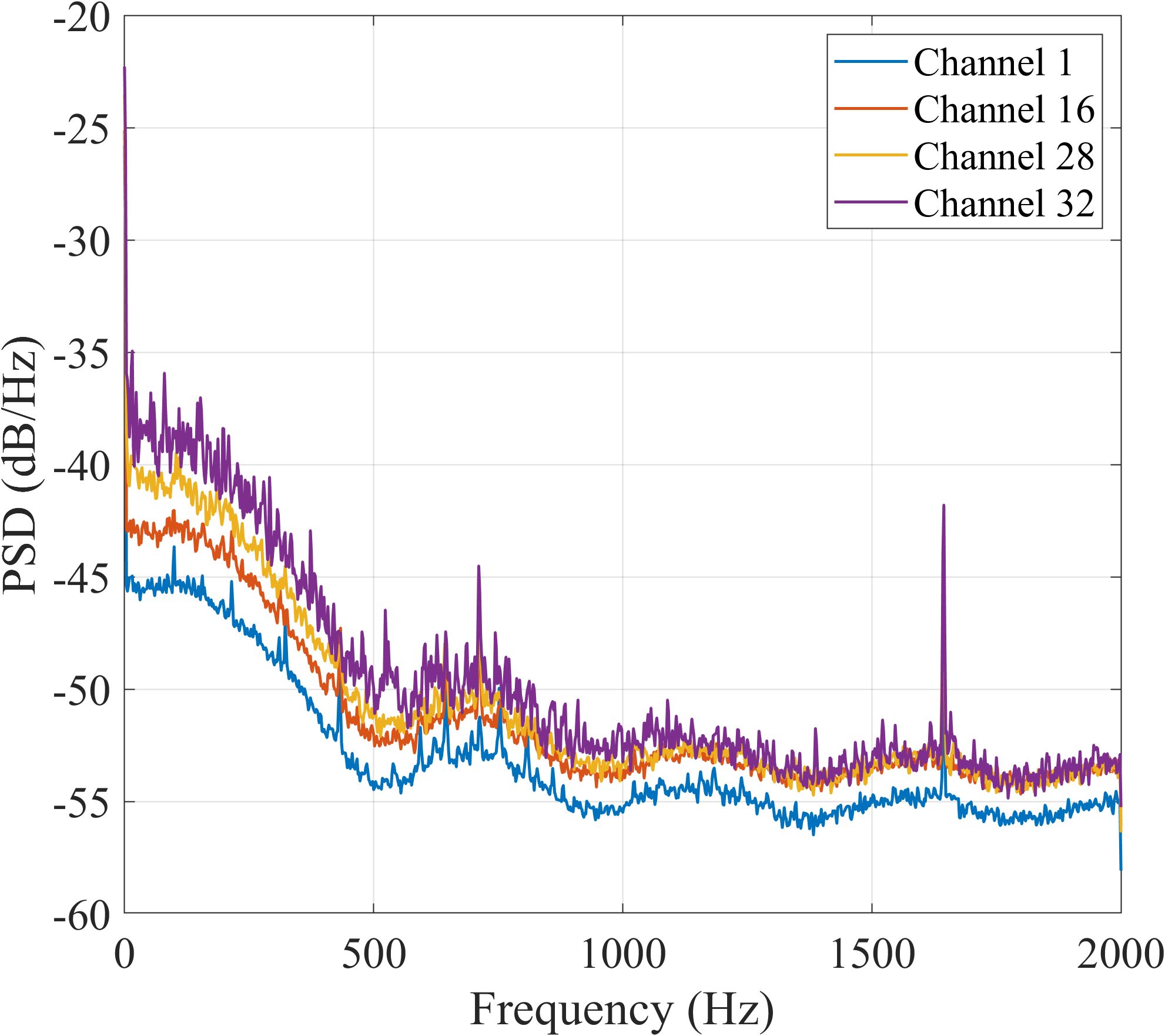}
	\caption{The power spectral density of ambient noise across DAS channels.}
	\label{Fig:PSD}
	\vspace{-3mm}
\end{figure}
\begin{figure*}[h]
	\centerline{\includegraphics[width=0.8\textwidth]{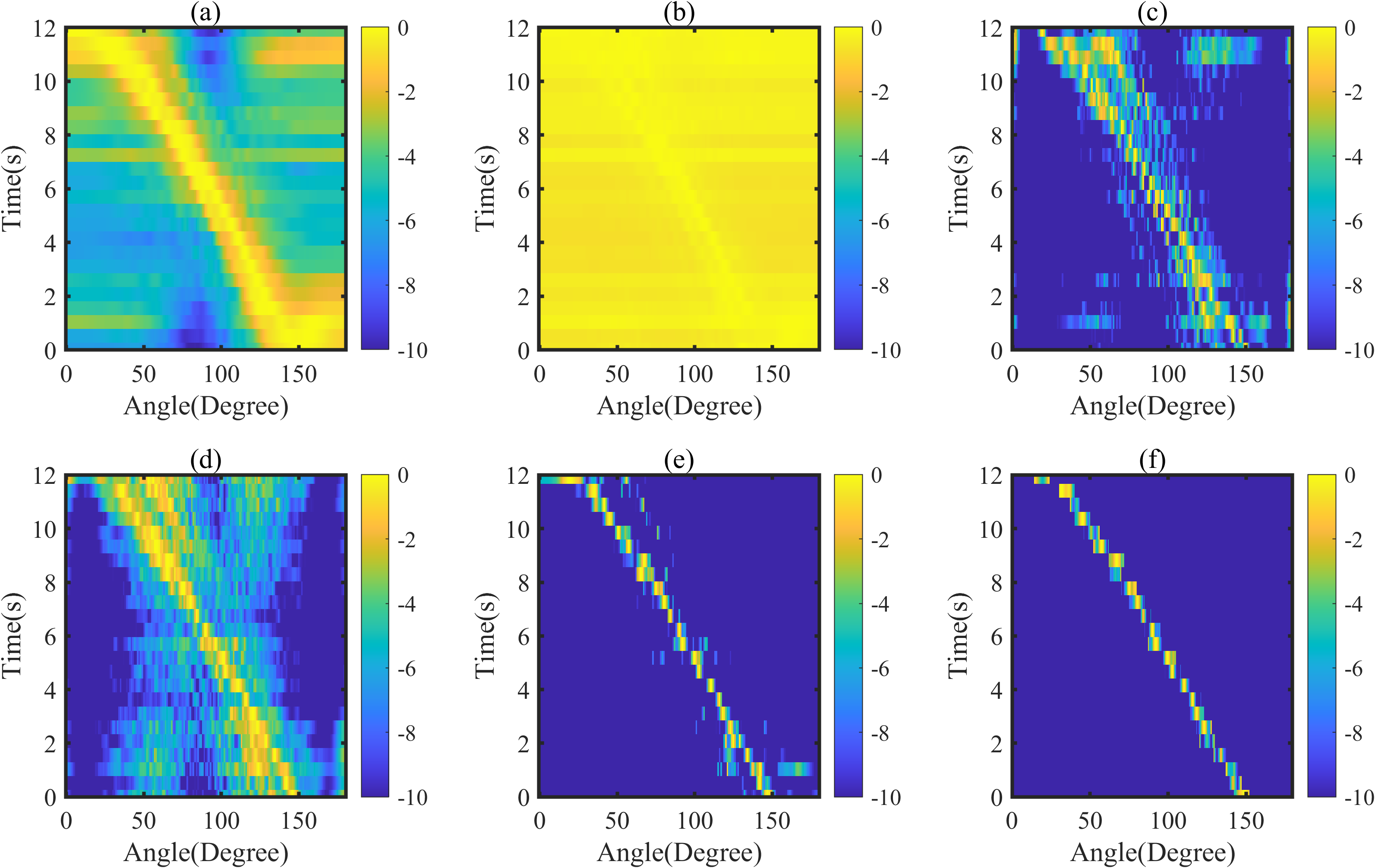}}
	\caption{The experimental data processing results involve the use of DAS data for tracking the speedboat. (a) CBF. (b) MUSIC. (c) SBL. (d) SPICE. (e) $q$-SPICE. (f) $q$-SPICE-$\text{GNR}^2$.}
	\label{Fig:ExprimTResult_fastBoat_BTR}
	\vspace{-3mm}
\end{figure*}

Source 2 initially emitted a sinusoidal wave at 900Hz, with a signal duration of 3 seconds and intervals of 2 seconds. Subsequently, the data was displayed after selecting the middle channel of the DAS array. This specific channel is regarded as being in the identical sound field as the standard hydrophone. The waveform and spectrogram of the signal collected in the selected channel are depicted in Fig.\ref{Fig:TankDASwaveform} and Fig.\ref{Fig:TankDASTF}. The waveform highlights the strong signal response capability of the DAS system using the spiral sensitized optical cable. The spectrogram shows that the center frequency of the signal aligns well with the frequency of the transmitted signal. 

The normalized waveform from the DAS system and a standard hydrophone is shown in Fig.\ref{Fig:CompareDAS}. The normalized cross-correlation coefficient at 900 Hz is 0.973. This analysis was also extended to other frequencies within the 700 Hz to 1200 Hz range, measured at 100 Hz intervals in an anechoic water tank. As shown in Fig.\ref{Fig:CC}, the correlation coefficients between the signals acquired by the DAS system and the standard hydrophone remain consistently high, ranging from 0.95 to 0.99 across all tested frequencies. These results demonstrate that the signal acquisition performance of the DAS system is highly comparable to that of the standard hydrophone.

Two sound sources simultaneously transmit signals. The inter-element spacing of the DAS array is set to 0.635 meters. The spiral-sensitized optical cable, which is situated underwater, covers a distance of roughly 5 meters and can accommodate 8 channels. The signals from these channels are processed for azimuth estimation using the $q$-SPICE framework. The bearing time recording results, shown in Fig.\ref{Fig:DAStankBTR}, highlight that the DAS system can achieve high-resolution azimuth estimation for dual sound sources, while maintaining a low background noise level.

\subsection{Lake Experiment}
We now address the challenge of DOA sensing in a non-uniform noise environment with complex topography using our own experimental data. The ambient noise within the reservoir exhibited non-uniform characteristics, primarily due to human activities such as fishing and tourism. A spiral-sensitized optical cable was deployed in the Xinfengjiang Reservoir for data acquisition. For target detection and bearing estimation, 32 channels were selected, with detailed data analysis focused on 10 representative channels. The corresponding waveforms are presented in Fig.\ref{Fig:32waveforms}. As shown in Fig.\ref{Fig:32waveforms}, the received signals exhibit varying levels of fluctuation across different channels, highlighting significant non-uniformity caused by occasional high-amplitude impulsive noise. The distribution of noise energy across channels, shown in Fig.\ref{Fig:noise}, corresponds closely with the observed waveform fluctuations. For power spectral analysis, representative data from four channels were selected, as illustrated in Fig.\ref{Fig:PSD}. The results reveal substantial variation in noise power among channels; for example, at 523 Hz, the noise power difference reaches approximately 8 dB. Furthermore, the lake bottom exhibits an irregular topography, a consequence of its history as a man-made reservoir containing numerous submerged structures from former settlements. This complex terrain results in multiple protrusions and gullies, causing deviations from a typical horizontal linear array configuration and introducing positional errors in sensor spacing.

As a passive monitoring system, it is capable of detecting numerous surrounding underwater sources. On June 5, 2023, a speedboat was observed passing near the DAS system. Radiation noise from the speedboat was recorded over a period of 12 seconds, and the corresponding data segment was extracted for subsequent analysis. For broadband DOA estimation for all methods, frequency points were selected at 10 Hz intervals within the 50–1050 Hz range, using an FFT length of 512. The grid configuration for the steering vector, which spans from 0° to 180°, was equally segmented into 181 grid points for CBF, MUSIC, SBL, SPICE, and $q$-SPICE. While $q$-SPICE-$\text{GNR}^2$, the initial grid was also set with a 1-degree interval. For grid redefinition, a 0.05-degree interval was used.

\begin{table}[!t]
	\centering
	\caption{Processing times of six DOA estimation approaches applied to the 12-second DAS array dataset. The third column reports the relative processing time, normalized with respect to the $q$-SPICE-$\text{GNR}^2$ method.}
	\label{tab:processing_times}
	\vspace{2pt}
	\begin{tabular}{lcc}
		\toprule
		Method & Processing time (s) & Relative runtime  \\
		\midrule
		CBF          & 13.1 & 0.14 \\
		MUSIC & 124.0 & 1.3 \\
		SBL & 1003.3 & 10.8 \\
	   SPICE & 2196.7 & 23.6 \\
		$q$-SPICE & 3199.8 & 34.4 \\
	   $q$-SPICE-$\text{GNR}^2$ & \textbf{93.0} & \textbf{1.0} \\
		\bottomrule
	\end{tabular}
\end{table}

Fig.~\ref{Fig:ExprimTResult_fastBoat_BTR} presents the BTR results, which indicate that both the CBF and MUSIC methods effectively captured the boat's trajectory. However, these methods exhibited limited resolution. Specifically, non-uniform noise caused an increase in the side lobes of the spatial power spectrum obtained using the MUSIC method, resulting in elevated background noise levels. Due to the non-cooperative nature of the target and the absence of ground truth trajectory data, the CBF results were used as a reference. Compared to CBF and MUSIC, the SPICE and SBL methods exhibit reduced background levels but generate false peaks at 10-12 seconds. In contrast, the $q$-SPICE and $q$-SPICE-$\text{GNR}^2$ exhibited the most stable performance in detection and tracking. The proposed $q$-SPICE-$\text{GNR}^2$ method outperformed the others in terms of DOA estimation resolution, background noise levels, and trajectory tracking continuity.

 To further evaluate the computational efficiency, we compared the runtime of six DOA estimation methods on the 12-second DAS dataset, using frequency-domain wideband processing. The evaluation was conducted on a laptop with an Intel Core i7-1260P (2.10 GHz) CPU and 32 GB DDR4 RAM. For a fair comparison, the steering vector grids of the baseline methods (CBF, MUSIC, SBL, SPICE, and $q$-SPICE) were uniformly discretized with a $0.1^\circ$ resolution over $[0^\circ,180^\circ]$, yielding 1801 grid points. In contrast, the proposed q-SPICE-GNR starts with a coarse grid ($1^\circ$ resolution, 181 points) and subsequently refines the dictionary through iterative updates, achieving the same final resolution of $0.1^\circ$. 
 
 The processing times are summarized in TABLE \ref{tab:processing_times}. As can be seen, CBF requires 13.1 s, MUSIC 124~s, and $q$-SPICE takes 3199.8~s, respectively. While the proposed $q$-SPICE-$\text{GNR}^2$ completes in 93~s. Although $q$-SPICE-$\text{GNR}^2$ involves iterative grid refinement, its staged strategy substantially reduces the computational burden, making it more efficient than SPICE and $q$-SPICE and significantly faster than SBL and MUSIC. The only faster alternative is CBF, but its azimuth resolution is notably lower, making it unsuitable for high-resolution DAS applications. These results confirm that the proposed iterative refinement not only maintains high-resolution capability but also delivers favorable runtime performance. 
\section{Conclusion} \label{Sec6}
In this work, we presented a novel underwater acoustic sensing system and target azimuth estimation approach by integrating a DAS system with a generalized sparse covariance-fitting framework in a broadband manner. By leveraging a custom-designed spiral-sensitized optical fiber cable, the DAS system demonstrated enhanced acoustic sensitivity, exhibiting high signal fidelity and strong correlation with conventional hydrophone measurements in controlled tank experiments.

Extensive simulations confirmed the advantages of the proposed $q$-SPICE-$\text{GNR}^2$ over conventional beamforming and existing sparse techniques, particularly in scenarios involving non-uniform noise and array imperfections.  The practical performance was further validated through field experiments conducted in a lake environment, which exhibited complex topography and diverse anthropogenic noise. Our DAS-based underwater acoustic sensing system reliably tracked a non-cooperative speedboat, while conventional methods exhibited degraded performance due to elevated side lobes and false detections.

Future research will concentrate on integrating learning-based models to enhance real-time target detection and classification. Overall, the integration of DAS technology with advanced sparse signal processing offers a promising pathway toward high-resolution, cost-effective, and scalable underwater acoustic sensing systems for applications in naval surveillance, marine ecology, and oceanographic research. Moreover, a more systematic investigation will be conducted under optimal low-noise conditions, using extended static measurements to thoroughly explore the correlation between system noise (including potential Brillouin and Raman scattering effects) and sensitivity.
\FloatBarrier
\bibliography{SIYUAN}

@Article{stoica2010spice,
  author    = {Stoica, Petre and Babu, Prabhu and Li, Jian},
  journal   = {IEEE Transactions on Signal Processing},
  title     = {SPICE: A sparse covariance-based estimation method for array processing},
  year      = {2010},
  number    = {2},
  pages     = {629--638},
  volume    = {59},
  publisher = {IEEE},
}

@Article{ahmed2021performance,
  author    = {Ahmed, Nauman and Wang, Huigang and Raja, Muhammad Asif Zahoor and Ali, Wasiq and Zaman, Fawad and Khan, Wasim Ullah and He, Yigang},
  journal   = {IEEE Access},
  title     = {Performance analysis of efficient computing techniques for direction of arrival estimation of underwater multi targets},
  year      = {2021},
  pages     = {33284--33298},
  volume    = {9},
  publisher = {IEEE},
}

@Article{liu2023robust,
  author    = {Liu, Aifei and Shi, Shengguo and Wang, Xinyi},
  journal   = {IEEE Transactions on Geoscience and Remote Sensing},
  title     = {Robust DOA estimation method for underwater acoustic vector sensor array in presence of ambient noise},
  year      = {2023},
  month     = jul,
  pages     = {4206014},
  volume    = {61},
  publisher = {IEEE},
}

@Article{liu2024grant,
  author    = {Liu, Ya-Feng and Yu, Wei and Wang, Ziyue and Chen, Zhilin and Sohrabi, Foad},
  journal   = {Next Generation Multiple Access},
  title     = {Grant-Free Random Access via Covariance-Based Approach},
  year      = {2024},
  pages     = {391--414},
  publisher = {Wiley Online Library},
}

@Article{liu2022robustVBI,
  author    = {Liu, Ying and Zhang, Zongyu and Zhou, Chengwei and Yan, Chenggang and Shi, Zhiguo},
  journal   = {IEEE Transactions on Vehicular Technology},
  title     = {Robust variational Bayesian inference for direction-of-arrival estimation with sparse array},
  year      = {2022},
  number    = {8},
  pages     = {8591--8602},
  volume    = {71},
  publisher = {IEEE},
}

@Article{Lim2016GMUSIC,
  author    = {Lim, Hock Siong and Ng, Boon Poh and Reddy, Vinod V},
  journal   = {IEEE Journal of Oceanic Engineering},
  title     = {Generalized MUSIC-like array processing for underwater environments},
  year      = {2016},
  number    = {1},
  pages     = {124--134},
  volume    = {42},
  publisher = {IEEE},
}

@Article{sward2018generalized,
  author    = {Sw{\"a}rd, Johan and Adalbj{\"o}rnsson, Stefan I and Jakobsson, Andreas},
  journal   = {Signal Processing},
  title     = {Generalized sparse covariance-based estimation},
  year      = {2018},
  pages     = {311--319},
  volume    = {143},
  publisher = {Elsevier},
}

@Article{Guo2023Variational,
  author    = {Guo, Kun and Zhang, Liang and Li, Yingsong and Zhou, Tian and Yin, Jingwei},
  journal   = {IEEE Transactions on Aerospace and Electronic Systems},
  title     = {Variational Bayesian inference for DOA estimation under impulsive noise and nonuniform noise},
  year      = {2023},
  number    = {5},
  pages     = {5778--5790},
  volume    = {59},
  publisher = {IEEE},
}

@Article{zhang2018wideband,
  author    = {Zhang, Yongchao and Jakobsson, Andreas and Zhang, Yin and Huang, Yulin and Yang, Jianyu},
  journal   = {IEEE Transactions on Geoscience and Remote Sensing},
  title     = {Wideband sparse reconstruction for scanning radar},
  year      = {2018},
  number    = {10},
  pages     = {6055--6068},
  volume    = {56},
  publisher = {IEEE},
}

@Article{schmidt1986multiple,
  author    = {Schmidt, Ralph},
  journal   = {IEEE transactions on antennas and propagation},
  title     = {Multiple emitter location and signal parameter estimation},
  year      = {1986},
  number    = {3},
  pages     = {276--280},
  volume    = {34},
  publisher = {IEEE},
}

@Article{vallet2015performance,
  author    = {Vallet, Pascal and Mestre, Xavier and Loubaton, Philippe},
  journal   = {IEEE Transactions on Signal Processing},
  title     = {Performance analysis of an improved MUSIC DoA estimator},
  year      = {2015},
  number    = {23},
  pages     = {6407--6422},
  volume    = {63},
  publisher = {IEEE},
}

@Article{luo2014challenges,
  author    = {Luo, Yu and Pu, Lina and Zuba, Michael and Peng, Zheng and Cui, Jun-Hong},
  journal   = {IEEE Transactions on Emerging Topics in Computing},
  title     = {Challenges and opportunities of underwater cognitive acoustic networks},
  year      = {2014},
  number    = {2},
  pages     = {198--211},
  volume    = {2},
  publisher = {IEEE},
}

@Article{pan2023underwater,
  author    = {Pan, Huaxin and Tang, Kangxu and Zhuo, Jia and Lu, Yuming and Chen, Jialong and Lv, Zhichao},
  journal   = {Journal of Marine Science and Engineering},
  title     = {Underwater acoustic technology-based monitoring of oil spill: A review},
  year      = {2023},
  number    = {4},
  pages     = {870},
  volume    = {11},
  publisher = {MDPI},
}

@Article{liu2024continuous,
  author    = {Liu, Li and Zhao, Tengfei and Chan, Sammy and Wu, Changmao},
  journal   = {IEEE Transactions on Mobile Computing},
  title     = {Continuous Object Tracking via Joint Global Local Binary Tree Topological Transformation in Underwater Acoustic Sensor Networks},
  year      = {2024},
  issn      = {1558-0660},
  month     = dec,
  number    = {12},
  pages     = {11091 - 11104},
  volume    = {23},
  publisher = {IEEE},
}

@Article{gerg2024deep,
  author    = {Gerg, Isaac D and Cook, Daniel A and Monga, Vishal},
  journal   = {IEEE Journal of Selected Topics in Applied Earth Observations and Remote Sensing},
  title     = {Deep Adaptive Phase Learning: Enhancing Synthetic Aperture Sonar Imagery Through Learned Coherent Autofocus},
  year      = {2024},
  month     = apr,
  pages     = {9517 - 9532},
  volume    = {17},
  publisher = {IEEE},
}

@Article{xu2019internet,
  author    = {Xu, Guobao and Shi, Yanjun and Sun, Xueyan and Shen, Weiming},
  journal   = {Sensors},
  title     = {Internet of things in marine environment monitoring: A review},
  year      = {2019},
  number    = {7},
  pages     = {1711},
  volume    = {19},
  publisher = {MDPI},
}

@Article{mikhalevsky2020deep,
  author    = {Mikhalevsky, Peter N and Sperry, Brian J and Woolfe, Katherine F and Dzieciuch, Matthew A and Worcester, Peter F},
  journal   = {The Journal of the Acoustical Society of America},
  title     = {Deep ocean long range underwater navigation},
  year      = {2020},
  number    = {4},
  pages     = {2365--2382},
  volume    = {147},
  publisher = {AIP Publishing},
}

@Article{chen2024method,
  author    = {Chen, Feng and Yang, Desen and Mo, Shiqi and Huang, Xingyu and Wang, Mingguang and Zhu, Zhongrui and Zhang, Haoyang},
  journal   = {IEEE Transactions on Geoscience and Remote Sensing},
  title     = {A method for estimating the direction of arrival without knowing the source number using acoustic vector sensor arrays},
  year      = {2024},
  month     = apr,
  pages     = {1001016},
  volume    = {62},
  publisher = {IEEE},
}

@Article{fu2019robust,
  author    = {Fu, Hua and Abeywickrama, Samith and Yuen, Chau and Zhang, Meng},
  journal   = {IEEE Transactions on Vehicular Technology},
  title     = {A robust phase-ambiguity-immune DOA estimation scheme for antenna array},
  year      = {2019},
  number    = {7},
  pages     = {6686--6696},
  volume    = {68},
  publisher = {IEEE},
}

@Article{yang2024sparse,
  author    = {Yang, Jianyu and Li, Wenchao and Li, Kefeng and Chen, Rui and Zhang, Kun and Mao, Deqing and Zhang, Yin},
  journal   = {IEEE Journal of Selected Topics in Applied Earth Observations and Remote Sensing},
  title     = {Sparse Bayesian Learning-based Multichannel Radar Forward-Looking Superresolution Imaging Considering Grid Mismatch},
  year      = {2024},
  month     = aug,
  pages     = {14997 - 15008},
  volume    = {17},
  publisher = {IEEE},
}

@Article{tuo2023radar,
  author    = {Tuo, Xingyu and Mao, Deqing and Zhang, Yin and Zhang, Yongchao and Huang, Yulin and Yang, Jianyu},
  journal   = {IEEE Journal of Selected Topics in Applied Earth Observations and Remote Sensing},
  title     = {Radar forward-looking super-resolution imaging using a two-step regularization strategy},
  year      = {2023},
  pages     = {4218--4231},
  volume    = {16},
  publisher = {IEEE},
}

@InProceedings{qiu2021directional,
  author       = {Qiu, Wei and Ma, Shuqing and Yan, Bing and Li, Le and Bao, Changchun and Zhang, Lilun},
  booktitle    = {2021 OES China Ocean Acoustics (COA)},
  title        = {Directional-of-Arrival Estimation for A Novel Non-uniform Linear Array},
  year         = {2021},
  organization = {IEEE},
  pages        = {846--849},
}

@Article{pesavento2023three,
  author    = {Pesavento, Marius and Trinh-Hoang, Minh and Viberg, Mats},
  journal   = {IEEE Signal Processing Magazine},
  title     = {Three more decades in array signal processing research: An optimization and structure exploitation perspective},
  year      = {2023},
  number    = {4},
  pages     = {92--106},
  volume    = {40},
  publisher = {IEEE},
}

@Article{Krim1996two,
  author    = {Krim, Hamid and Viberg, Mats},
  journal   = {IEEE signal processing magazine},
  title     = {Two decades of array signal processing research: the parametric approach},
  year      = {1996},
  number    = {4},
  pages     = {67--94},
  volume    = {13},
  publisher = {IEEE},
}

@Article{ghafoor2019overview,
  author    = {Ghafoor, Huma and Noh, Youngtae},
  journal   = {IEEE Access},
  title     = {An overview of next-generation underwater target detection and tracking: An integrated underwater architecture},
  year      = {2019},
  pages     = {98841--98853},
  volume    = {7},
  publisher = {IEEE},
}

@Article{lu2024underwater,
  author    = {Lu, Yi and Yuan, Yufan and Liu, Meiyan and Wei, Yan and Tu, Xingbin and Qu, Fengzhong},
  journal   = {IEEE Sensors Journal},
  title     = {An Underwater Wideband Sound Source Localization Method Based on Light Neural Network Structure},
  year      = {2024},
  month     = jul,
  number    = {13},
  pages     = {20970 - 20980},
  volume    = {24},
  publisher = {IEEE},
}

@Article{Men2024Joint,
  author    = {Men, Wei and Du, Jun and Yin, Jingwei and Zhang, Liang and Liu, Lei and Ren, Yong and Niyato, Dusit},
  journal   = {IEEE Transactions on Wireless Communications},
  title     = {Joint Detection and Communication System Design via Combination of Index and Phase Modulations},
  year      = {2024},
  month     = oct,
  pages     = {15690 - 15704},
  volume    = {23},
  publisher = {IEEE},
}

@Article{gerstoft2016multisnapshot,
  author    = {Gerstoft, Peter and Mecklenbr{\"a}uker, Christoph F and Xenaki, Angeliki and Nannuru, Santosh},
  journal   = {IEEE Signal Processing Letters},
  title     = {Multisnapshot sparse Bayesian learning for DOA},
  year      = {2016},
  number    = {10},
  pages     = {1469-1473},
  volume    = {23},
  publisher = {IEEE},
}

@Article{malioutov2005sparse,
  author    = {Malioutov, Dmitry and Cetin, M{\"u}jdat and Willsky, Alan S},
  journal   = {IEEE transactions on signal processing},
  title     = {A sparse signal reconstruction perspective for source localization with sensor arrays},
  year      = {2005},
  number    = {8},
  pages     = {3010--3022},
  volume    = {53},
  publisher = {IEEE},
}

@Article{zhang2021online,
  author    = {Zhang, Yongchao and Li, Jie and Li, Minghui and Zhang, Yin and Luo, Jiawei and Huang, Yulin and Yang, Jianyu and Jakobsson, Andreas},
  journal   = {IEEE Geoscience and Remote Sensing Letters},
  title     = {Online sparse reconstruction for scanning radar using beam-updating q-SPICE},
  year      = {2021},
  month     = feb,
  pages     = {1-5},
  volume    = {19},
  publisher = {IEEE},
}

@Article{lu2021distributed,
  author    = {Lu, Bin and Wu, Bingyan and Gu, Jinfeng and Yang, Junqi and Gao, Kan and Wang, Zhaoyong and Ye, Lei and Ye, Qing and Qu, Ronghui and Chen, Xiaobao and others},
  journal   = {Optics express},
  title     = {Distributed optical fiber hydrophone based on $\Phi$-OTDR and its field test},
  year      = {2021},
  number    = {3},
  pages     = {3147--3162},
  volume    = {29},
  publisher = {Optical Society of America},
}

@Article{bouffaut2022eavesdropping,
  author    = {Bouffaut, L{\'e}a and Taweesintananon, Kittinat and Kriesell, Hannah J and R{\o}rstadbotnen, Robin A and Potter, John R and Landr{\o}, Martin and Johansen, St{\aa}le E and Brenne, Jan K and Haukanes, Aksel and Schjelderup, Olaf and others},
  journal   = {Frontiers in Marine Science},
  title     = {Eavesdropping at the speed of light: Distributed acoustic sensing of baleen whales in the Arctic},
  year      = {2022},
  pages     = {901348},
  volume    = {9},
  publisher = {Frontiers Media SA},
}

@Article{landro2022sensing,
  author    = {Landr{\o}, Martin and Bouffaut, L{\'e}a and Kriesell, Hannah Joy and Potter, John Robert and R{\o}rstadbotnen, Robin Andr{\'e} and Taweesintananon, Kittinat and Johansen, St{\aa}le Emil and Brenne, Jan Kristoffer and Haukanes, Aksel and Schjelderup, Olaf and others},
  journal   = {Scientific Reports},
  title     = {Sensing whales, storms, ships and earthquakes using an Arctic fibre optic cable},
  year      = {2022},
  number    = {1},
  pages     = {19226},
  volume    = {12},
  publisher = {Nature Publishing Group UK London},
}

@Article{cang2025deploying,
  author    = {Cang, Siyuan and Xu, Min and Chen, Jiantong and Li, Chao and Gao, Kan and Jiang, Xingda and Wang, Zhaoyong and Luo, Bin and Xiao, Zhuo and Guo, Zhen and others},
  journal   = {Journal of Marine Science and Engineering},
  title     = {Deploying an Integrated Fiber Optic Sensing System for Seismo-Acoustic Monitoring: A Two-Year Continuous Field Trial in Xinfengjiang},
  year      = {2025},
  number    = {2},
  pages     = {368},
  volume    = {13},
  publisher = {MDPI},
}

@Article{liu2024voiceprint,
  author    = {Liu, Yifan and Li, Chao and Chen, Yici and Wang, Zhaoyong and Wu, Bingyan and Wu, Jinyi and Chen, Boqi and Yang, Huayong and Gao, Kan and Ye, Qing and others},
  journal   = {Optics Express},
  title     = {Voiceprint and position information detection of non-cooperative ship with $\Phi$-OTDR and suspended sensitized optical cable},
  year      = {2024},
  number    = {10},
  pages     = {17362--17372},
  volume    = {32},
  publisher = {Optica Publishing Group},
}

@Article{shao2025tracking,
  author    = {Shao, Jie and Wang, Yibo and Zhang, Yixin and Zhang, Xuping and Zhang, Chi},
  journal   = {IEEE Geoscience and Remote Sensing Letters},
  title     = {Tracking Moving Ships Using Distributed Acoustic Sensing Data},
  year      = {2025},
  month     = jan,
  pages     = {7502605},
  volume    = {22},
  publisher = {IEEE},
}

@Article{chen2025near,
  author    = {Chen, Junfeng and Li, Hao and Ai, Ke and Zhu, Shuolong and Fan, Cunzheng and Yan, Zhijun and Sun, Qizhen},
  journal   = {IEEE Transactions on Instrumentation and Measurement},
  title     = {Near-field Multi-source Localization and Signal Enhancement with ASP-assisted DAS},
  year      = {2025},
  month     = feb,
  number    = {9504311},
  volume    = {74},
  publisher = {IEEE},
}

@Article{matsumoto2021detection,
  author    = {Matsumoto, Hiroyuki and Araki, Eiichiro and Kimura, Toshinori and Fujie, Gou and Shiraishi, Kazuya and Tonegawa, Takashi and Obana, Koichiro and Arai, Ryuta and Kaiho, Yuka and Nakamura, Yasuyuki and others},
  journal   = {Scientific reports},
  title     = {Detection of hydroacoustic signals on a fiber-optic submarine cable},
  year      = {2021},
  number    = {1},
  pages     = {2797},
  volume    = {11},
  publisher = {Nature Publishing Group UK London},
}

@Article{douglass2023distributed,
  author    = {Douglass, Alexander S and Abadi, Shima and Lipovsky, Bradley P},
  journal   = {JASA Express Letters},
  title     = {Distributed acoustic sensing for detecting near surface hydroacoustic signals},
  year      = {2023},
  number    = {6},
  volume    = {3},
  publisher = {AIP Publishing},
}

@Article{wilcock2023distributed,
  author    = {Wilcock, William SD and Abadi, Shima and Lipovsky, Bradley P},
  journal   = {JASA Express Letters},
  title     = {Distributed acoustic sensing recordings of low-frequency whale calls and ship noise offshore Central Oregon},
  year      = {2023},
  number    = {2},
  volume    = {3},
  publisher = {AIP Publishing},
}

@Article{lan2020acoustical,
  author    = {Lan, Hualin and Lv, Yunfei and Jin, Jianjia and Li, Jianghui and Sun, Dajun and Yang, Zhiguo},
  journal   = {IEEE Internet of Things Journal},
  title     = {Acoustical observation with multiple wave gliders for internet of underwater things},
  year      = {2020},
  number    = {4},
  pages     = {2814--2825},
  volume    = {8},
  publisher = {IEEE},
}

@Article{baron2020hydrophone,
  author    = {Baron, Valentin and Finez, Arthur and Bouley, Simon and Fayet, Florent and Mars, J{\'e}r{\^o}me I and Nicolas, Barbara},
  journal   = {IEEE Journal of Oceanic Engineering},
  title     = {Hydrophone array optimization, conception, and validation for localization of acoustic sources in deep-sea mining},
  year      = {2020},
  number    = {2},
  pages     = {555--563},
  volume    = {46},
  publisher = {IEEE},
}

@Article{hu2020decentralized,
  author    = {Hu, Li and Wang, Xiaodong and Wang, Shilian},
  journal   = {IEEE Sensors Journal},
  title     = {Decentralized underwater target detection and localization},
  year      = {2020},
  number    = {2},
  pages     = {2385--2399},
  volume    = {21},
  publisher = {IEEE},
}

@Article{he2021optical,
  author    = {He, Zuyuan and Liu, Qingwen},
  journal   = {Journal of Lightwave Technology},
  title     = {Optical fiber distributed acoustic sensors: A review},
  year      = {2021},
  number    = {12},
  pages     = {3671--3686},
  volume    = {39},
  publisher = {IEEE},
}

@Article{heshmati2020robust,
  author    = {Heshmati-Alamdari, Shahab and Nikou, Alexandros and Dimarogonas, Dimos V},
  journal   = {IEEE Transactions on Automation Science and Engineering},
  title     = {Robust trajectory tracking control for underactuated autonomous underwater vehicles in uncertain environments},
  year      = {2021},
  month     = jul,
  number    = {3},
  pages     = {1288--1301},
  volume    = {18},
  publisher = {IEEE},
}

@Article{colosi2020observations,
  author    = {Colosi, John A and Rudnick, Daniel L},
  journal   = {The Journal of the Acoustical Society of America},
  title     = {Observations of upper ocean sound-speed structures in the North Pacific and their effects on long-range acoustic propagation at low and mid-frequencies},
  year      = {2020},
  number    = {4},
  pages     = {2040--2060},
  volume    = {148},
  publisher = {AIP Publishing},
}

@Article{jiajing2019distributed,
  author    = {Jiajing, Liang and Zhaoyong, Wang and Bin, Lu and Xiao, Wang and Luchuan, Li and Qing, Ye and Ronghui, Qu and Haiwen, Cai},
  journal   = {Optics letters},
  title     = {Distributed acoustic sensing for 2D and 3D acoustic source localization},
  year      = {2019},
  number    = {7},
  pages     = {1690--1693},
  volume    = {44},
  publisher = {Optical Society of America},
}

@Article{guang2024sensitivity,
  author    = {Guang, Dong and Sun, Xiaoyuan and Shi, Jinhui and Wu, Xuqiang and Zhang, Guosheng and Zuo, Cheng and Zhu, Pengcheng and Yu, Benli},
  journal   = {Optics Express},
  title     = {Sensitivity improvement of fiber optic interferometric hydrophone based on composite structure},
  year      = {2024},
  number    = {27},
  pages     = {47721--47734},
  volume    = {32},
  publisher = {Optica Publishing Group},
}

@Article{wang2021large,
  author    = {Wang, Ting and Zhao, Xiaodong and Du, Hongliang and Xia, Song and Li, Guo and Guo, Haisheng and Li, Fei and Xu, Zhuo},
  journal   = {IEEE Transactions on Ultrasonics, Ferroelectrics, and Frequency Control},
  title     = {Large-area piezoelectric single crystal composites via 3-D-printing-assisted dice-and-insert technology for hydrophone applications},
  year      = {2021},
  number    = {10},
  pages     = {3241--3248},
  volume    = {68},
  publisher = {IEEE},
}

@Article{fouda2025phase,
  author    = {Fouda, Bernard Marie Tabi and Wand, Lei and Zhang, Wenjun and Atangana, Jacques},
  journal   = {IEEE Transactions on Instrumentation and Measurement},
  title     = {Phase-Sensitive Optical Time-Domain Reflectometry-Based Audio Excitation Signal Demodulation and Reproduction},
  year      = {2025},
  month     = jan,
  number    = {7001414},
  volume    = {74},
  publisher = {IEEE},
}

@Article{zhou2015distributed,
  author    = {Zhou, Ling and Wang, Feng and Wang, Xiangchuan and Pan, Yun and Sun, Zhenqin and Hua, Ji and Zhang, Xuping},
  journal   = {IEEE Photonics Technology Letters},
  title     = {Distributed strain and vibration sensing system based on phase-sensitive OTDR},
  year      = {2015},
  number    = {17},
  pages     = {1884--1887},
  volume    = {27},
  publisher = {IEEE},
}

@Article{li2014fiber,
  author    = {Li, Qin and Zhang, Chunxi and Li, Chuansheng},
  journal   = {Optik},
  title     = {Fiber-optic distributed sensor based on phase-sensitive OTDR and wavelet packet transform for multiple disturbances location},
  year      = {2014},
  number    = {24},
  pages     = {7235--7238},
  volume    = {125},
  publisher = {Elsevier},
}

@Article{shao2019optimal,
  author    = {Shao, Guoliang and Guo, Yong-Xin},
  journal   = {IEEE Transactions on Instrumentation and Measurement},
  title     = {An optimal design for passive magnetic localization system based on SNR evaluation},
  year      = {2019},
  number    = {7},
  pages     = {4324--4333},
  volume    = {69},
  publisher = {IEEE},
}
\bibliographystyle{IEEEtran}

\end{document}